\documentclass[aps,prl,nofootinbib,superscriptaddress,twocolumn,backend=biber,floatfix,letterpaper,url=false,backend=biber]{revtex4}
\usepackage[colorlinks,urlcolor=blue,citecolor=blue,linkcolor=blue]{hyperref}
\usepackage{graphicx}
\usepackage{epstopdf}
\usepackage{amsmath,amsthm,amssymb,latexsym,amsfonts,times,mathrsfs,verbatim}
\usepackage{mathptmx}
\usepackage{braket}
\usepackage{lineno}
\usepackage{bigints}
\usepackage[usenames,dvipsnames]{color}

\DeclareMathAlphabet{\mathcal}{OMS}{cmsy}{m}{n}

\newcommand\etal{{\em et al.}}
\def\ie{\emph{i.e.}}
\def\eg{\emph{e.g.}}

\def\omm{\omega_{\rm m}} 
\def\ommt{\tilde{\omega}_{\rm m}} 
\def\nb{\bar{n}_{\rm b}} 
\def\Gamth{\Gamma_{\rm th}} 
\def\nth{\bar{n}_{\rm th}} 

\def\omc{\omega_{\rm c}} 
\def\omp{\omega_{\rm p}} 
\def\omd{\omega_{\rm d}} 
\def\Delp{\Delta_{\rm p}} 
\def\Delc{\Delta_{\rm c}} 
\def\epsp{\epsilon_{\rm p}^{\phantom{*}}} 
\def\epsd{\epsilon_{\rm d}^{\phantom{*}}} 
\def\epsps{\epsilon_{\rm p}^*} 
\def\epsds{\epsilon_{\rm d}^*} 
\def\na{\bar{n}_{\rm a}} 
\def\np{\bar{n}_{\rm p}} 

\def\go{g^{\phantom{*}}_1} 
\def\gos{g^*_1} 
\def\gt{g^{\phantom{*}}_2} 
\def\gts{g^*_2} 
\def\liou{\mathcal{L}} 
\def\epst{\epsilon^{\phantom{*}}_2} 
\def\epsts{\epsilon^*_2} 
\def\Gamlin{\Gamma_{\rm lin}} 
\def\Gamex{\Gamma_{\rm ex}} 
\def\Gamdec{\Gamma_{\rm dec}} 

\def\W+{\mathcal{W}_+} 

\def\kB{k_{\rm B}} 

\newcommand\Tr{\mathrm{Tr}}

\newcommand\red{}
\newcommand\redtwo{}

\begin{document}

\title{Nonlinear Sideband Cooling to a Cat State of Motion}
\author{B.D. Hauer}\email{bradley.hauer@nist.gov}
\affiliation{National Institute of Standards and Technology, Boulder, Colorado 80305, USA}
\author{J. Combes}
\affiliation{Department of Electrical, Computer, and Energy Engineering, University of Colorado, Boulder, Colorado 80309, USA}
\author{J.D. Teufel}\email{john.teufel@nist.gov}
\affiliation{National Institute of Standards and Technology, Boulder, Colorado 80305, USA}

\date{\today}

\begin{abstract}

The ability to prepare a macroscopic mechanical resonator into a quantum superposition state is an outstanding goal of cavity optomechanics. Here, we propose a technique to generate cat states of motion using the intrinsic nonlinearity of a dispersive optomechanical interaction. By applying a bichromatic drive to an optomechanical cavity, our protocol enhances the inherent second-order processes of the system, inducing the requisite two-phonon dissipation. We show that this nonlinear sideband cooling technique can dissipatively engineer a mechanical resonator into a cat state, which we verify using the full Hamiltonian and an adiabatically reduced model. While the fidelity of the cat state is maximized in the single-photon, strong-coupling regime, we demonstrate that Wigner negativity persists even for weak coupling. Finally, we show that our cat state generation protocol is robust to significant thermal decoherence of the mechanical mode, indicating that such a procedure may be feasible for near-term experimental systems.

\end{abstract}

\maketitle

{\it Introduction}---Engineered micro/nanomechanical systems have recently emerged as viable resources for quantum technologies \cite{barzanjeh_2022} and fundamental tests of quantum mechanics \cite{chen_2013}. At the forefront of this effort is cavity optomechanics \cite{aspelmeyer_2014}, which utilizes sideband techniques to stabilize quantum states of motion via reservoir engineering \cite{poyatos_1996}. Canonical examples include ground state cooling \cite{teufel_2011b, chan_2011}, squeezing \cite{wollman_2015, lecocq_2015}, and entanglement \cite{palomaki_2013, riedinger_2018, ockeloenkorppi_2018, kotler_2021} of mechanical motion. However, these protocols rely on a strong pump to parametrically enhance the linear coupling between the cavity field and mechanical motion, obscuring the inherent nonlinearity of the interaction. Thus, the current field of quantum optomechanics is largely confined to performing bilinear operations on Gaussian states. By reinforcing the intrinsic optomechanical nonlinearity, one could break free of this Gaussian prison and prepare interesting nonclassical states of mechanical motion. 

Of particular interest are macroscopic superposition states known as cat states \cite{schrodinger_1935}, which have previously been observed in trapped ions \cite{munroe_1996}, confined photons \cite{brune_1996}, and superconducting circuits \cite{leghtas_2015, lescanne_2020}. Though recent experiments have prepared non-Gaussian states of mechanical motion using nonlinearities derived from superconducting qubits \cite{oconnell_2010, chu_2018, ma_2021, wollack_2022} and single-photon detection \cite{riedinger_2018, patel_2021, enzian_2021}, the experimental generation of macroscopic superposition states has yet to be demonstrated. Preparing these highly nonclassical states in mechanical resonators would allow them to be used as quantum-enhanced sensors \cite{munro_2002, joo_2011, jacobs_2017}, nodes in quantum communication networks \cite{sangouard_2010, brask_2010}, long-lived, error-protected qubits \cite{leghtas_2013, chamberland_2022}, and platforms to study macroscopic quantum collapse theories \cite{bose_1999}.

Early proposals to create mechanical cat states utilized the intrinsic Kerr nonlinearity of the optomechanical interaction \cite{bose_1997, bose_1999}, however, this method requires vacuum coupling larger than both the mechanical frequency and cavity loss rate. Subsequent optomechanical cat state generation protocols have suggested introducing nonlinearities via single-photon detection to perform conditional measurements \cite{marshall_2003, akram_2013} or single-phonon addition/subtraction \cite{zhan_2020, shomroni_2020}, coupling with external two-level systems \cite{pflanzer_2013, yin_2013}, swaps between nonclassical cavity states and mechanical modes \cite{hoff_2016, teh_2018}, or using time-varying electromagnetic fields \cite{ge_2015, liao_2016a, brawley_2016, davis_2018, clarke_2019}. Each of these proposals rely on some combination of probabilistic measurements, complicated integration with external sources of nonlinearity {\redtwo (\eg~qubits)}, and/or generation of complex electromagnetic fields. Beyond being difficult to implement experimentally, the complexity of these methods will incur inefficiencies in the mechanical cat state preparation, decreasing its fidelity. On the other hand, optomechanical protocols that utilize reservoir engineering techniques \cite{poyatos_1996} have recently been proposed for the deterministic and stable generation of macroscopic mechanical superposition states \cite{tan_2013a, asjad_2014, brunelli_2018, brunelli_2019}. Such schemes utilize continuous, coherent electromagnetic sources and are robust to external decoherence. Unfortunately, these proposed methods have relied on quadratic coupling between the cavity and mechanical element, which is small relative to its destructive linear counterpart \cite{thompson_2008, doolin_2014}. 

{\redtwo Here, we introduce a reservoir engineering technique that uses the nonlinearity inherent to all optomechanical systems to prepare cat states of motion. This simple scheme differentiates itself from previous reservoir engineering proposals, as it circumvents the experimentally prohibitive requirement for direct coupling to the square of mechanical displacement. Furthermore, our technique obviates the need for external sources of nonlinearity or complex electromagnetic drives, providing a resource-efficient method to generate macroscopic superposition states.} For this protocol, one applies two continuous-wave pumps to an optomechanical cavity, one resonant and one red-detuned by twice the mechanical frequency. This prepares the mechanical resonator into a cat state via two-phonon sideband cooling and squeezing processes. Using a master equation approach, we deduce that mechanical cat state generation is feasible for vacuum coupling rates that are greater than the cavity loss rate, but much smaller than the mechanical frequency. By numerically simulating the full optomechanical Hamiltonian, we verify that our protocol can generate high-fidelity mechanical cat states with near ideal Wigner negativity. Finally, we show that our protocol is robust to contamination from the surrounding environment, allowing for preparation of mechanical superposition states in the presence of thermal noise.

\begin{figure}[t!]
\centerline{\includegraphics[width=\columnwidth]{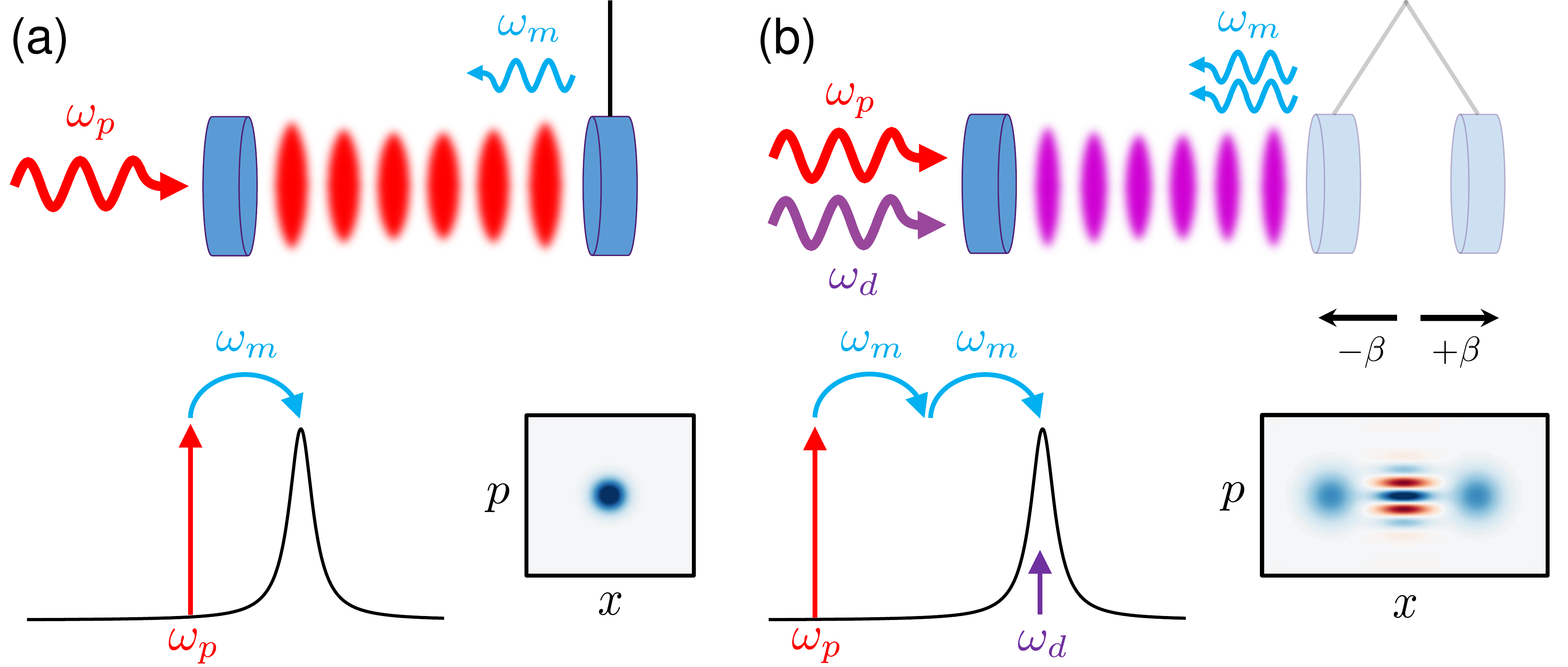}}
\caption{{\label{fig1}} Schematic, frequency-space, and phase-space representations of a) conventional optomechanical sideband cooling and b) our proposed method of mechanical cat state generation using nonlinear sideband cooling.}
\end{figure}

{\it Dissipation engineering protocol.}---Our procedure is adapted from a previous technique used to prepare cat states in superconducting microwave cavities coupled via a Josephson nonlinearity \cite{leghtas_2015, lescanne_2020}. In this protocol, the cat state is prepared in the long-lived ``storage'' cavity, while a second ``readout'' cavity is used to engineer nonlinear dissipation processes. This protocol naturally maps to an optomechanical system, whereby the mechanical resonator becomes the ``storage'' element, with the coupled electromagnetic cavity providing the fast ``readout.'' The requisite nonlinearity is then provided by the optomechanical interaction itself in place of a Josephson junction. With this architecture, one can apply a strong pump to the cavity red-detuned by twice the mechanical frequency to mediate two-phonon cooling processes (see Fig.~\ref{fig1}). A second tone applied on resonance will then mix with this detuned pump providing a two-phonon mechanical drive. Combined, these two-phonon processes are parity preserving, such that a resonator initialized into an even (odd) state will be restricted to the even (odd) manifold of its Fock basis and will evolve into an even (odd) cat state \cite{tan_2013a}. Along with these nonlinear processes, linear optomechanical coupling will persist in any realistic system, acting to flip the parity of the desired cat state and contribute to its decoherence \cite{kim_1992}. Therefore, we must treat the optomechanical Hamiltonian in its entirety, such that both linear and nonlinear terms are included.

To investigate our protocol, we begin with the optomechanical master equation
\begin{equation}
\label{optomast}
\dot{\rho} = -\frac{i}{\hbar} [H,\rho] + \kappa L[a] \rho + \Gamma (\nb + 1) L[b] \rho + \Gamma \nb L[b^\dag] \rho,
\end{equation}
where $a$ ($b$) is the annihilation operator for the electromagnetic cavity (mechanical resonator). Here, we assume a zero temperature bath for the cavity (\ie~$\bar{n}_a = 0$) with total loss rate $\kappa$ and frequency $\omc$. Meanwhile, the mechanical mode with frequency $\omm$ is thermalized at its decay rate $\Gamma$ to an environment at finite temperature $T$ and average occupancy $\nb = \left( e^{\hbar \omm / \kB T} - 1 \right)^{-1}$. We have also introduced the Lindblad superoperator $L[o] \rho = o \rho o^\dag - \frac{1}{2} o^\dag o \rho - \frac{1}{2} \rho o^\dag o$ for arbitrary operator $o$ and density matrix $\rho$, as well as the optomechanical Hamiltonian in the frame rotating at the pump frequency $\omp$
\begin{equation}
\begin{split}
\label{dispHam}
\frac{H}{\hbar} = &-\Delta a^\dag a + \omm b^\dag b + \epsd a^\dag e^{i \Delp t} + \epsds a e^{-i \Delp t} \\
&+ (\gos a + \go a^\dag) (b + b^\dag) + g_0 a^\dag a (b + b^\dag).
\end{split}
\end{equation}
Here, we have translated the cavity mode by its steady state amplitude $\alpha = \epsp / (\Delta + i \kappa /2)$ in the presence of a coherent pump with amplitude $\epsp$, where $\Delta = \omp - \omc$, such that $a$ now acts on the displaced cavity mode. The average number of photons in the cavity due to this pump can be calculated as $\np = |\alpha|^2$. {\red Note that in Eq.~\eqref{dispHam} we have implicitly accounted for the static displacement in mechanical equilibrium position due to this steady state photon population, as well as the corresponding shift in cavity frequency, by appropriately modifying our reference frame \cite{SM}.} Also included is a coherent cavity drive with amplitude $\epsd$ and frequency $\omd$ detuned from the pump frequency by $\Delp = \omp - \omd$. Finally, the last two terms in Eq.~\eqref{dispHam} characterize the linearized and nonlinear optomechanical interactions, with the linear cavity-enhanced coupling rate $\go = \alpha g_0$ expressed in terms of the optomechanical vacuum coupling rate $g_0$.  \nocite{dodonov_1974, glauber_1963, wigner_1932, gerry_1997, kenfack_2004, grimm_2020, mirrahimi_2014, luttinger_1955, schrieffer_1966, nakajima_1958, zwanzig_1960, zwanzig_1964, lorch_2015, gautier_2022}

While Eq.~\eqref{dispHam} gives an exact description of our optomechanical cavity, often the final term is neglected, leading to a linearized description. However, this term is crucial in providing the nonlinearity required for our protocol. We therefore use a Schrieffer-Wolff transformation \cite{machado_2016} with the generator $S = \frac{g_0}{\omm} a^\dag a ( b^\dag - b)$ \cite{nunnenkamp_2011} to expand this term to leading order in $g_0 / \omm$, which amounts to the replacement \cite{SM}
\begin{equation}
\begin{split}
\label{nlomexp}
& g_0 a^\dag a (b + b^\dag) \Rightarrow \left( \gts a - \gt a^\dag \right) \left(b^2 - b^\dag{}^2 \right) \\
&- \frac{g_0^2}{\omm} \left\{ \left(a^\dag a \right)^2 + \alpha^* \left(2 a^\dag a + 1 \right) a + \alpha a^\dag \left( 2 a^\dag a + 1 \right)  \right\}.
\end{split}
\end{equation}
The first term in Eq.~\eqref{nlomexp} elucidates the fact that the intrinsic optomechanical nonlinearity can be used to mediate second-order processes whereby single photons simultaneously interact with two phonons at a strength determined by the second-order coupling rate $\gt = g_0^2 \alpha / \omm$. Meanwhile the second term corresponds to the higher-order corrections to the electromagnetic cavity in the presence of optomechanical coupling, including the well-known self-Kerr nonlinearity \cite{aspelmeyer_2014}.

{\it Adiabatically eliminated model.}---To proceed, we assume that the cavity's decay rate $\kappa$ {\red exceeds the interaction rates of the mechanical resonator with its thermal bath and the electromagnetic cavity \cite{SM}.} In this regime, the system will quickly equilibrate to its steady state, which we take to be the slow subspace spanned by the cavity's ground state and the full Hilbert space of the mechanical mode. We then adiabatically eliminate the cavity mode by tracing over its rapidly evolving excited states. This is performed using the Nakajima-Zwanzig formalism \cite{wilsonrae_2008} to derive the reduced master equation for the mechanical mode in the sideband-resolved regime ($\kappa \ll \omm$) as
\begin{align}
\begin{split}
\label{MEb}
\dot{\rho}_b &= - \frac{i}{\hbar} [H_b, \rho_b] + \Gamma_2 L[b^2] \rho_b + \Gamlin L[b] \rho_b + \Gamex L[b^\dag] \rho_b,
\end{split}
\end{align}
where
\begin{align}
\label{Hb}
\frac{H_b}{\hbar} = \epst b^\dag{}^2 + \epsts b^2 - K \left( b^\dag b \right)^2.
\end{align}
Here, we have chosen to maximize cat state generation efficiency by applying the drive tone on resonance with the cavity ($\omd = \omc$) and the pump tone at $\omp = \omc - 2 \ommt$, where $\ommt$ is the dressed mechanical frequency \cite{SM}.

\begin{figure}[t!]
\centerline{\includegraphics[width=\columnwidth]{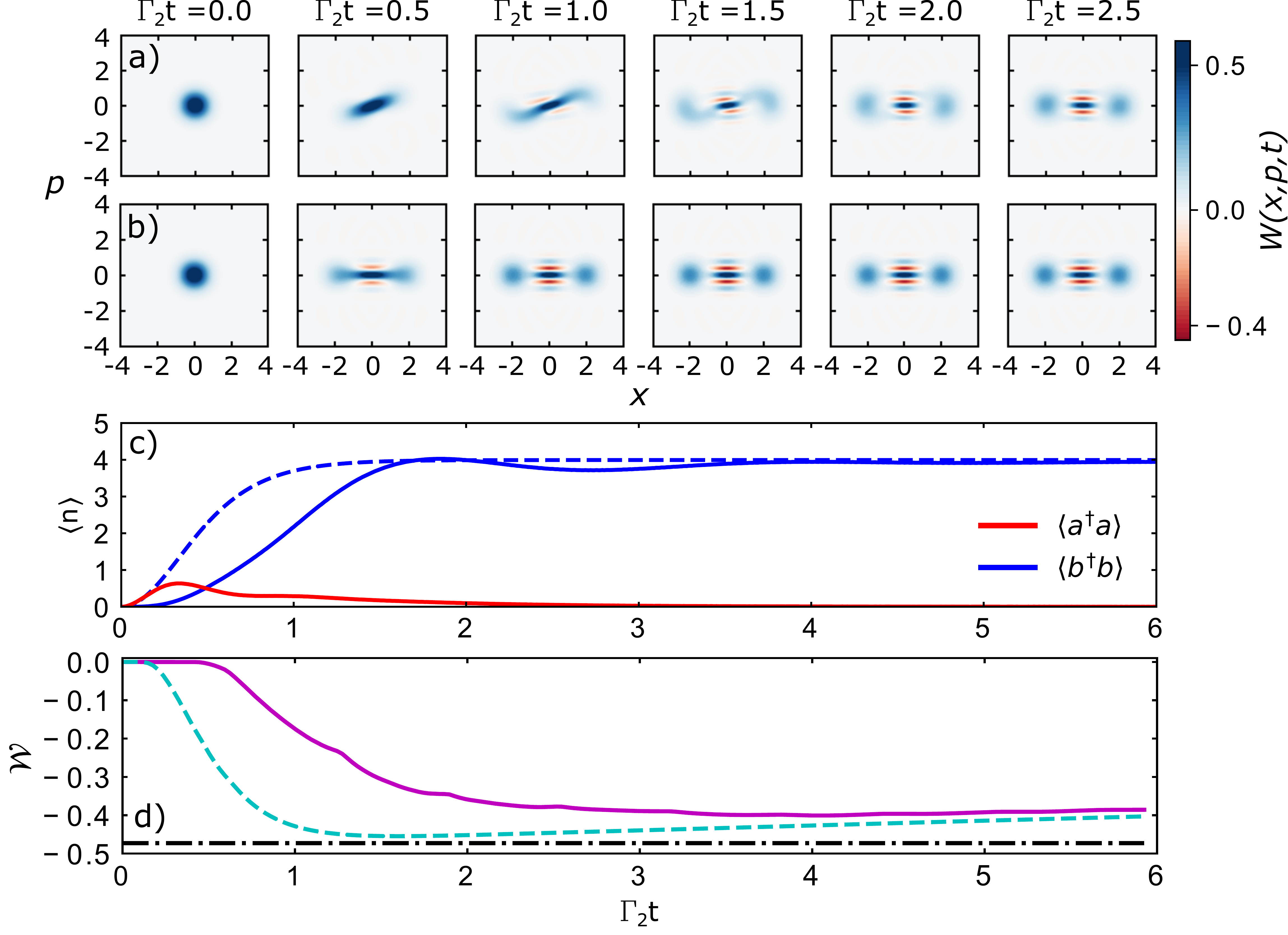}}
\caption{{\label{fig2}} Evolution of a mechanical resonator's Wigner distribution from its ground state to a $\beta = 2$ even cat state using our reservoir engineering protocol simulated with a) the full master equation in Eq.~\eqref{optomast} and b) the reduced master equation in Eq.~\eqref{MEb}. We also show c) the average occupancy of the mechanical and cavity modes and d) the minimal Wigner negativity of the mechanical state vs evolution time (full model - solid, reduced model - dashed). Simulation parameters are given in the main text. Here, we find that for the full (reduced) model the Wigner negativity is minimized to $\mathcal{W}_{\rm min} = -0.401$ ($\mathcal{W}_{\rm min} = -0.455$) at $\Gamma_2 t \approx$ 3.9 ($\Gamma_2 t \approx$ 1.6). This corresponds to a maximal fidelity of 95.6\% (99.1\%) with an ideal even cat state of the same size, whose minimal Wigner negativity ($\W+ \approx -0.476$) is indicated by the black dashed-dotted line in d).}
\end{figure}

Equation \eqref{MEb} includes all of the terms necessary to dissipatively engineer mechanical cat states. The first term describes the evolution of the resonator according to $H_b$. This Hamiltonian contains coherent two-phonon squeezing terms, with amplitude $\epsilon_2 = 2i \epsd \gts / \kappa$, that arise from the mixing of the pump tones applied to the cavity, along with a phonon-dependent Kerr nonlinearity with strength $K = |\gt|^2 / 4 \omm$. The second term describes two-phonon optomechanical cooling at a rate $\Gamma_2 = 4 |\gt|^2 / \kappa$. This engineered cooling, coupled with the squeezing terms, restricts the mechanical resonator to two-phonon operations, thus preserving its parity. Acting under these two processes alone, the resonator will naturally evolve from the ground state into an even cat state on a timescale set by $\tau \sim 1 / \Gamma_2$ (see Fig.~\ref{fig2}). The size of the resultant cat state is given by $\beta = \sqrt{|\epsd|/|\gt|}$, while its rotation in phase space is set by the relative phase between $\epsp$ and $\epsd$. 

Meanwhile, the last two terms in Eq.~\eqref{MEb} represent incoherent single-phonon loss and excitation processes. In the first case, decoherence is caused by the mechanical resonator emitting phonons at a rate $\Gamlin = \Gamth + \Gamma_1$ into either its intrinsic environmental bath at its thermal decoherence rate $\Gamth = (\nb + 1) \Gamma$, or to an optomechanically generated reservoir at rate $\Gamma_1 = |\go|^2 \kappa / \omm^2$. The term proportional to $L[b^\dag] \rho_b$ then corresponds to incoherent phononic excitations from these two baths at a rate $\Gamex = \nb \Gamma + \Gamma_1 / 9$. Both of these single-phonon processes will flip the parity of the cat state from even to odd (or vice versa), causing it to decohere at a rate $\Gamdec = 2|\beta|^2 \left( \Gamlin + \Gamex \right)$ \cite{kim_1992}. To generate mechanical cat states we require $\Gamlin \gg \Gamex$, so here we will focus on $\Gamlin$. However, we retain the incoherent excitation terms in our numerical analysis, as they will have noticeable effects on the cat state's coherence. Simulations using Eq.~\eqref{MEb} can be seen in Fig.~\ref{fig2}b, where we show the evolution of the mechanical resonator from its ground state into an even cat state of size $\beta=2$.

In the ideal situation where $\Gamth \ll \Gamma_1$, only optomechanically induced losses need be considered, which sets a fundamental limit on our protocol's ability to generate mechanical cat states as $\Gamma_2 \gtrsim \Gamma_1$, or equivalently,
\begin{align}
\label{catcond}
g_0 \gtrsim \frac{\kappa}{2}.
\end{align}
That is, the protocol outlined in this paper is optimal when the vacuum coupling rate of the system is on the order of, or exceeds, the loss rate of the cavity (\ie~approaching or residing in the single-photon, strong-coupling regime). We also note that for $\beta \ge 1$, Eq.~\eqref{catcond} ensures that the autonomously stabilized cat states generated by our protocol can be used in error correction protocols where single quanta loss is the dominant error channel \cite{lescanne_2020}.

\begin{figure}[t!]
\centerline{\includegraphics[width=\columnwidth]{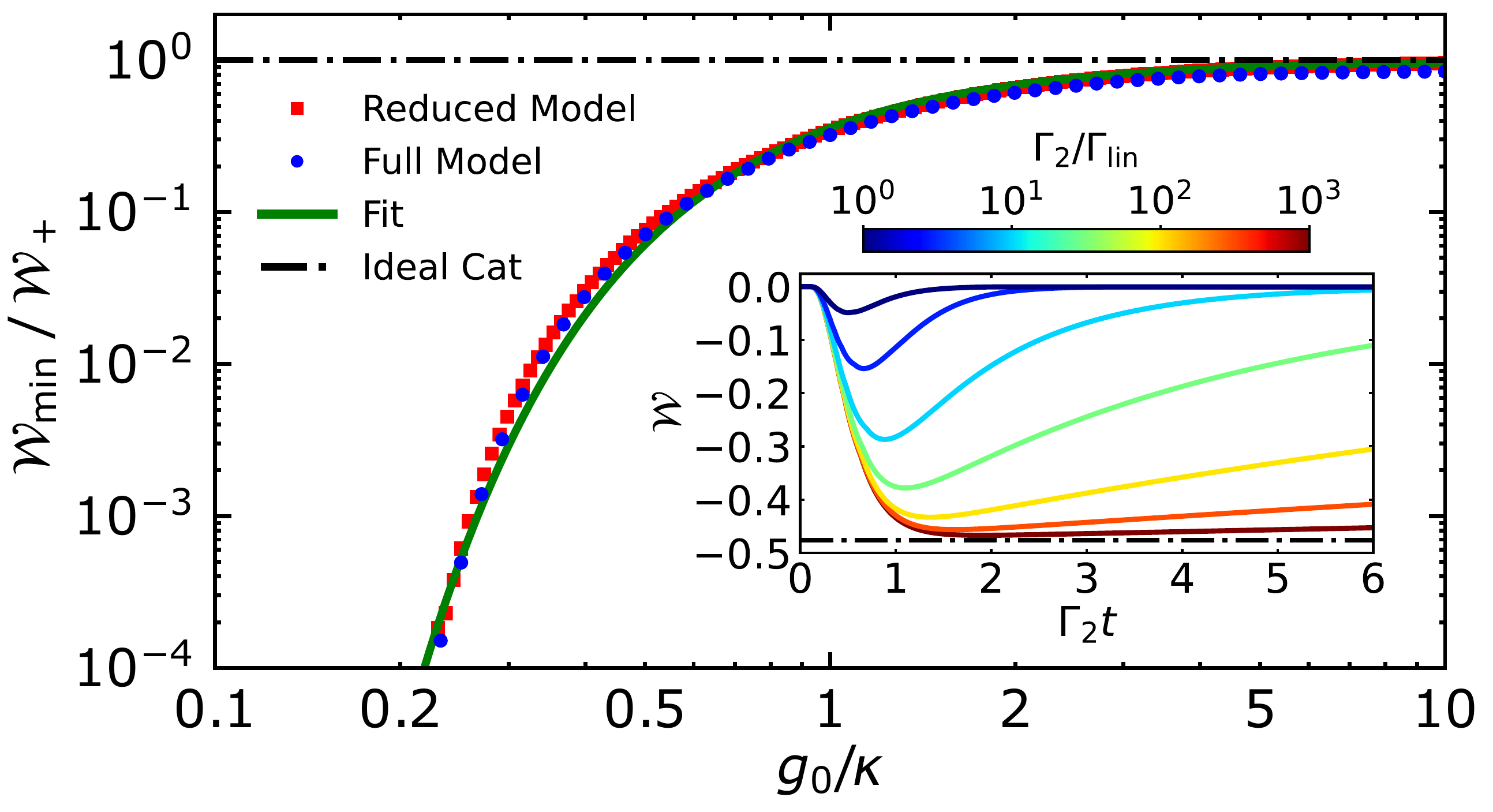}}
\caption{{\label{fig3}} $\mathcal{W}_{\rm min}$ (normalized by $\W+ = 0.476$) vs $g_0 / \kappa$ for the same parameters as Fig.~\ref{fig2} with varying $\kappa$ {\red and $\beta = 2$}. Here, we observe excellent agreement between the full (blue circles) and reduced (red squares) optomechanical models over nearly 2 orders of magnitude in $g_0 / \kappa$. Furthermore, the Wigner negativity approaches $\W+$ (black dashed-dotted line) for $g_0 / \kappa \gg 1/2$, while exponentially decreasing for $g_0 / \kappa < 1/2$ as expected. Also included is a fit of Eq.~\eqref{Wnegmintg0k} to the full model (solid green line). Inset: $\mathcal{W}$ vs time for varying ratios of $\Gamma_2 / \Gamlin$. Here, we have chosen $\kappa / 2 \pi$ = 10 kHz, such that $g_0 / \kappa = 100$, while varying $\Gamma$.}
\end{figure}

{\it Wigner negativity.}---To rigorously characterize the quantum superposition states generated using our procedure, we use the Wigner function $W(x,p)$, whose minimum we label the Wigner negativity or $\mathcal{W} = {\rm min}\{W(x,p)\}$ \cite{SM}. For dynamic systems, $\mathcal{W}$ can be further minimized in time, resulting in the minimal Wigner negativity $\mathcal{W}_{\rm min}$. Using a phenomenological model, we show that for $\Gamth \ll \Gamma_1$ the minimal Wigner negativity for our dissipation engineering protocol takes the simple form \cite{SM}
\begin{equation}
\label{Wnegmintg0k}
\mathcal{W}_{\rm min} = \W+ \exp \left[ -2 C |\beta|^2 \left( \frac{\kappa}{2 g_0} \right)^{2k+2} \right].
\end{equation}
Here, $C$ and $k$ are parameters that characterize the time required to reach this minimal Wigner negativity as
\begin{equation}
\label{tmin}
t_{\rm min} = \frac{C}{\Gamma_2} \left( \frac{\Gamma_1}{\Gamma_2} \right)^k,
\end{equation}
while $\W+$ is the Wigner negativity of an ideal even cat state. From Eq.~\eqref{Wnegmintg0k}, one can see that while Eq.~\eqref{catcond} sets the scale for mechanical cat state generation, it is a soft limit, as nonzero Wigner negativity persists for $2 g_0 < \kappa$. However, far below this limit the Wigner negativity dies off exponentially as a function of $\kappa / 2 g_0$ (see Fig.~\ref{fig3}).

{\it Numerical simulations.}---Using a numerical approach, we verify our model given in Eq.~\eqref{MEb} against the full optomechanical model of Eq.~\eqref{optomast}. This is shown in Fig.~\ref{fig2}, where we compare the evolution of a mechanical system from its ground state to a cat state using both models. Simulations are performed with the shared parameters $g_0 / 2 \pi$ = 1 MHz, $\omm / 2 \pi$ = 15 MHz, $\Gamma / 2 \pi$ = 15 Hz,  $\kappa / 2 \pi$ = 100 kHz, $\nb = 0$, and $\np$ = 0.1, with $\epsd$ chosen such that $\beta = 2$. This parameter set allows for fast simulation, and hence a concrete comparison between these two models. Though they differ at earlier times, both models agree well on long timescales where $\braket{a^\dag a} \approx 0$ and the cat state has stabilized, thus validating our adiabatically eliminated model.

We have also assessed how the condition in Eq.~\eqref{catcond} affects the minimal Wigner negativity of our generated cat states. This is presented in Fig.~\ref{fig3} using the same parameters as Fig.~\ref{fig2}, while changing $\kappa$ to vary the ratio $g_0 / \kappa$. We find that Eq.~\eqref{Wnegmintg0k} provides an excellent fit of the exact numerical results over multiple orders of magnitude in $g_0 / \kappa$ with the parameters $C \approx 1/3$ and $k \approx -1/4$. This indicates that the minimal Wigner negativity of the $\beta = 2$ cat state studied here decays exponentially in proportion to {\red $|\beta|^2 \left( \kappa / 2 g_0 \right)^{3/2}$}. In spite of this exponential behavior, significant Wigner negativity still persists for $g_0 / \kappa \lesssim 0.3$. Conversely, when $g_0 \gg \kappa$, the minimal Wigner negativities extracted from both models asymptotically approach their ideal values and high-fidelity cat states are created. We have additionally investigated the effect of increasing the mechanical decoherence rate $\Gamlin$ on the time dependence of the cat state's Wigner negativity (inset of Fig.~\ref{fig3}). Here, we confirm that the Wigner negativity is minimized on a timescale set by $1 / \Gamma_2$, with a weak dependence on $\Gamma_1 / \Gamma_2$ as indicated by Eq.~\eqref{tmin}. We further observe the minimal Wigner negativity approaches $\W+$ for $\Gamma_2 / \Gamlin \gg 1$, while decohering back to zero on a timescale set by $1 / \Gamlin$.

\begin{figure}[t!]
\centerline{\includegraphics[width=\columnwidth]{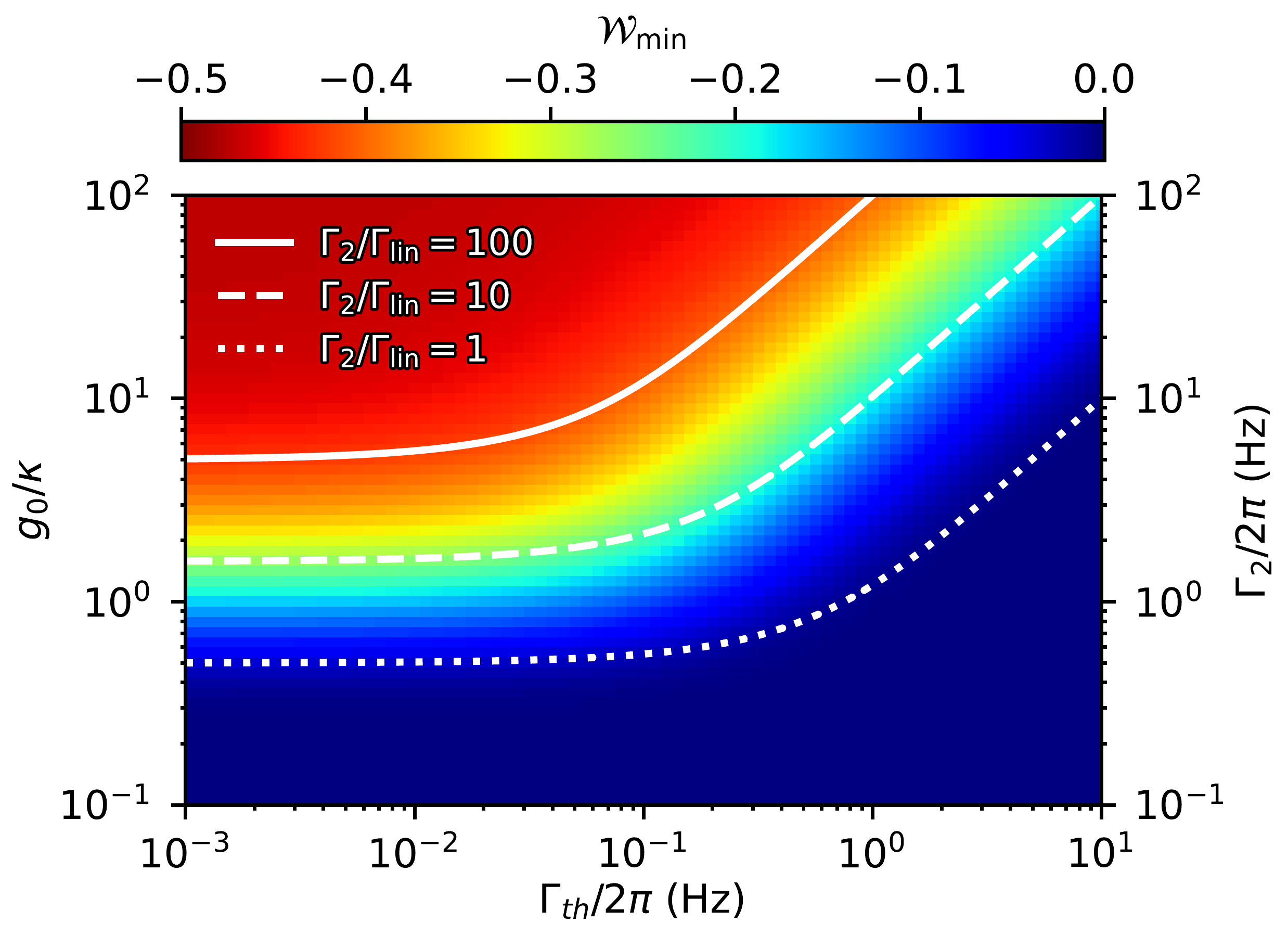}}
\caption{{\label{fig4}} Plot of $\mathcal{W}_{\rm min}$ vs $g_0 / \kappa$ and $\Gamth$ calculated using the reduced model in Eq.~\eqref{MEb}. Here, we have fixed $g_0 / 2 \pi$ = 10 kHz, {\red $\omm / 2 \pi$ = 20 MHz, $\bar{n}_b$ = 10 (corresponding to $T \approx$ 10 mK), and $\bar{n}_p$ = 100}, while $\kappa$ and $\Gamma$ are swept to vary $g_0 / \kappa$ (or equivalently $\Gamma_2$) and $\Gamth$, respectively. We have also chosen $\epsd$ such that the generated cat state is of size $\beta = 2$ at every point in the plot.}
\end{figure}

Finally, we have used our adiabatically eliminated model to investigate how $\mathcal{W}_{\rm min}$ varies over a large parameter space in $g_0 / \kappa$ and $\Gamth$, which is illustrated in Fig.~\ref{fig4}. On the right side of the plot where $\Gamlin \approx \Gamth$, regions of significant Wigner negativity are delineated by contours that exhibit a linear dependence between $g_0 / \kappa$ and $\Gamth$. Meanwhile on the left side, these boundaries plateau to a constant value. This is because the decoherence is no longer dominated by the thermal environment, but instead by pump-induced linear dissipation.  As $\Gamma_1$ and $\Gamma_2$ scale with pump power in the same way, their ratio is constant and given by $(2 g_0/\kappa)^2$ \cite{SM}. These results show clearly demarcated regions of parameters space that allow for significant Wigner negativity even in the presence of significant thermal noise.

{\it Experimental realization.}---Currently, the most challenging aspect of experimentally realizing our protocol is the condition given by Eq.~\eqref{catcond}. {\red While existing experiments in microwave circuits \cite{teufel_2011a} and optomechanical crystals \cite{chan_2012} have demonstrated $2 g_0 / \kappa \approx 0.01$, ongoing improvements in both coupling and cavity losses make each of these platforms a viable candidate for achieving $2 g_0 / \kappa \approx 1$. Specifically, in optomechanical crystals, numerical optimization of ultrasmall mode volume cavities projects to increase the vacuum coupling, while ensuring that radiation and fabrication uniformity do not limit the cavity losses \cite{bozkurt_2022}. In the microwave regime, bulk superconducting cavities already achieve sufficiently low loss \cite{reagor_2013} and ongoing advances in materials and surface preparation could allow similar quality factors in vacuum gap capacitor optomechanical circuits.  In addition to improvements in loss, recent proposals suggest that by pushing circuit optomechanics into the millimeter-wave regime, $g_0$ could also be increased by over an order of magnitude \cite{hauer_2021}. Together these innovations in increased coupling and cavity quality factor would enable experimental implementation of this cat state protocol.} Regardless of the platform, verification of these delicate quantum superposition states will require the ability to perform mechanical state tomography with low added noise.  Conveniently, cavity optomechanical systems have demonstrated nearly noiseless mechanical quadrature measurement techniques, using either quantum nondemolition \cite{lecocq_2015} or transient amplification \cite{delaney_2019} methods, which are sufficient for witnessing Wigner negativity.

{\it Conclusion.}---We have introduced a simple scheme that utilizes two continuous-wave pumps to cool an optomechanical resonator into a cat state of motion. To generate significant Wigner negativity with such a protocol, one must approach the single-photon, strong-coupling regime. Though unattainable in current experiments, with future improvements to state-of-the-art optomechanical systems, mechanical cat state generation using this protocol could soon be realized. This advancement would allow for straightforward preparation of long-lived mechanical cat states to be used as robust, rotation-symmetric bosonic codes for quantum computing \cite{grimsmo_2020}, or as canonical systems to study the fundamental collapse mechanisms of macroscopic quantum superposition states \cite{bose_1999}.

\begin{acknowledgments}

The authors would like to thank Matthew Woolley for fruitful discussions, as well as Scott Glancy and Nicholas Frattini for their careful reading of the manuscript and insightful comments. J.~C. was supported in part by NSF QLCI Grant No. OMA-2016244. B.~D.~H. acknowledges the support of the Natural Sciences and Engineering Research Council of Canada (NSERC) through NSERC (PDF-532793-2019) and Banting (BPF-180161) Postdoctoral Fellowships. Contributions to this article by workers at the National Institute of Standards and Technology, an agency of the U.S. Government, are not not subject to U.S. copyright.

\end{acknowledgments}

\bibliography{refs}

\clearpage
\widetext
\begin{center}
\textbf{\large Supplementary Information for Nonlinear Sideband Cooling to a Schr\"odinger Cat State of Motion}
\end{center}

\setcounter{equation}{0}
\setcounter{figure}{0}
\setcounter{table}{0}
\setcounter{section}{0}
\setcounter{page}{1}

\makeatletter

\renewcommand{\thefigure}{S\arabic{figure}}
\renewcommand{\theequation}{S\arabic{equation}}
\renewcommand{\thetable}{S\arabic{table}}
\renewcommand*{\citenumfont}[1]{S#1}
\renewcommand*{\bibnumfmt}[1]{[S#1]}

\section{Cat States}
\label{cat}

\subsection{General Properties}

For the purpose of this paper, we define cat states of size $\beta$ as the even and odd superpositions of two coherent states with complex amplitude $\beta$ and opposing phases. Mathematically, these states can be written as \cite{dodonov_1974_si}
\begin{equation}
\label{catstate}
\ket{\pm} = \frac{\ket{\beta} \pm \ket{-\beta}}{\sqrt{2 \left(1 \pm e^{-2 |\beta|^2} \right)}},
\end{equation}
where the plus (minus) sign corresponds to the even (odd) cat state and the coherent state $\ket{\beta}$ can be represented in the Fock basis $\ket{n}$ as \cite{glauber_1963_si}
\begin{equation}
\label{cohstate}
\ket{\beta} = e^{-\frac{|\beta|^2}{2}} \sum^\infty_{n=0} \frac{\beta^n}{\sqrt{n!}} \ket{n}.
\end{equation}
The probability distribution of the cat state in the Fock basis is then given by
\begin{equation}
\label{catprob}
P_n =|\braket{n | \pm}|^2 = \frac{e^{-|\beta|^2} \beta^{2n} \left[1 \pm (-1)^n \right]^2}{2 n! \left(1 \pm e^{-2 |\beta|^2} \right)},
\end{equation}
where $P_n$ is the probability of a measurement collapsing the cat state into the $n$th Fock state. From this probability distribution, one can immediately see that even (odd) cats are restricted to the even (odd) Fock state manifold of the resonator (see Fig.~\ref{figs1}). Using Eq.~\eqref{catstate} one can also show that $\braket{- |+} = \braket{+|-} = 0$ (\ie~the odd and even cat states span a two-state orthonormal basis) and that the average occupancy of a given cat state is $\bar{n}_\pm = \bra{\pm} b^\dag b \ket{\pm} = |\beta|^2$. Finally, we note that when $\beta$ is large ($|\beta| \gtrsim 2$), cat states are often approximated as $\ket{\pm} \approx \left( \ket{\beta} \pm \ket{-\beta} \right) / \sqrt{2}$. 

\subsection{Wigner Function}

\begin{figure}[h!]
\centerline{\includegraphics[width=4in]{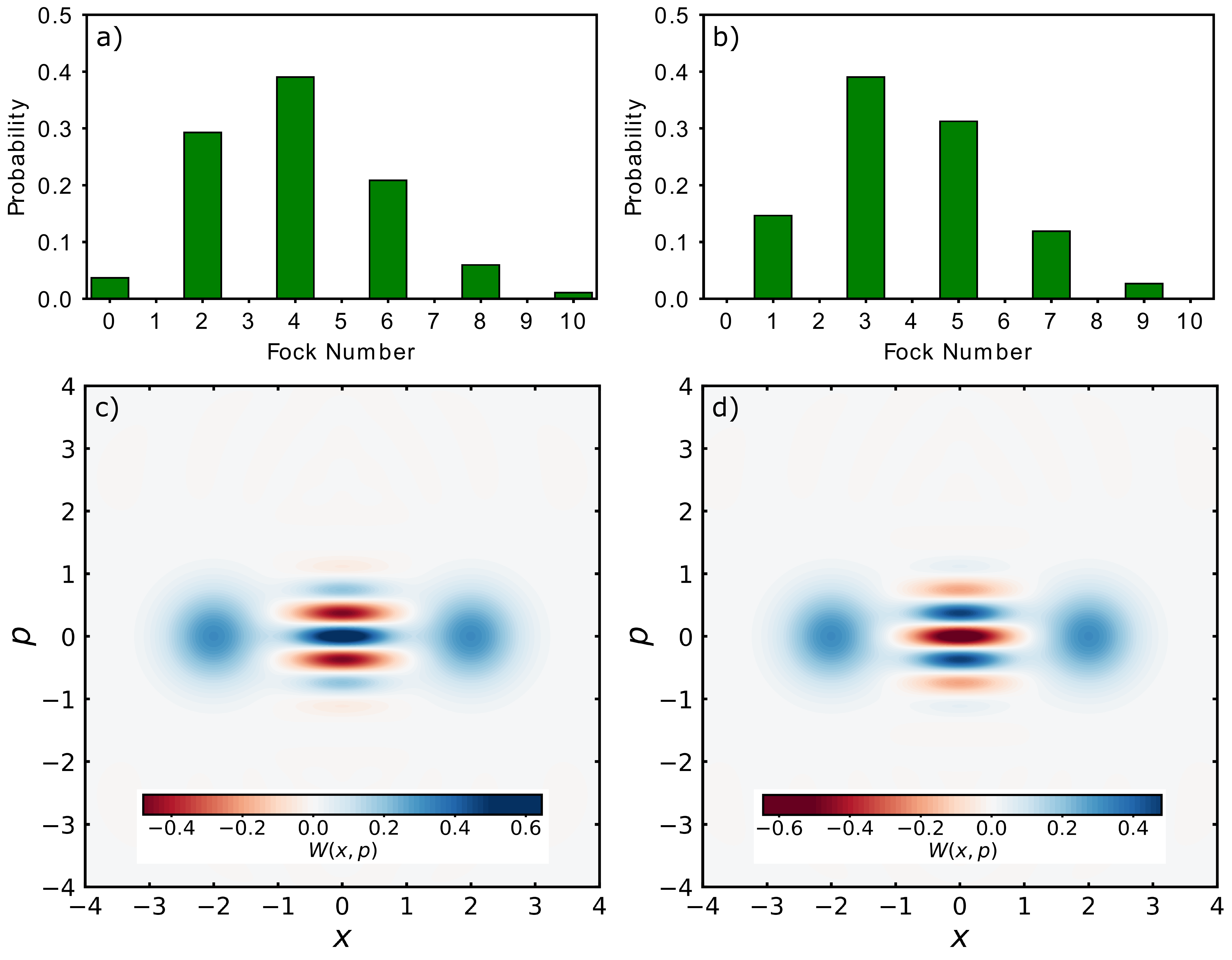}}
\caption{{\label{figs1}} Probability distribution of an a) even and b) odd cat state of size $\beta = 2$ in the Fock state basis according to Eq.~\eqref{catprob}. The Wigner functions calculated using Eq.~\eqref{Wigdef} are also shown for an c) even and d) odd cat state of the same size as those in a) and b).}
\end{figure}

A convenient way to visualize a cat state is through its two-dimensional Wigner quasiprobability distribution (also known as the Wigner function) \cite{wigner_1932_si}, which is defined in terms of the generalized conjugate position $x$ and momentum $p$ as
\begin{equation}
\label{Wigdef}
W(x,p) = \frac{1}{ \pi \hbar} \int_{-\infty}^{\infty} \bra{x+z} \rho \ket{x-z} e^{-2i pz / \hbar} dz,
\end{equation}
where $\rho$ is the density matrix of the system in question. Inputting the even and odd cat state vectors from Eq.~\eqref{catstate} into this expression, one finds the Wigner function for a pure cat state of size $\beta$ as \cite{gerry_1997_si}
\begin{equation}
\label{Wigcat}
W_{\pm}(x,p) = \frac{2 e^{-2 ( x^2 + p^2) }}{ \pi \left(1 \pm e^{-2 |\beta|^2} \right)} \left\{ \cosh(4 \beta x) e^{-2 | \beta|^2} \pm \cos(4 \beta p) \right\}.
\end{equation}
Fig.~\ref{figs1} gives a visualization of this distribution for both odd and even cat states.

One common measure of the nonclassicality of a quantum state is its Wigner negativity $\mathcal{W}$ \cite{kenfack_2004_si}, which we define here as the minimum value of the Wigner function throughout its entire phase space. This value is minimized for odd cat states at the origin and is given by
\begin{equation}
\label{Wignego}
\mathcal{W}_- \equiv {\rm min}\{W_-(x,p)\} = W_-(0,0) = -\frac{2}{\pi} \approx -0.637,
\end{equation}
such that the Wigner negativity of an odd cat is independent of its size. The situation is more complicated for even cat states, whose Wigner negativity is minimized for $x=0$ (for the phase chosen in Fig.~\ref{figs1}c), but depends on the size of the cat and can be found as
\begin{equation}
\label{Wignege}
\W+ \equiv {\rm min}\{W_+(x,p)\} = {\rm min}\{W_+(0,p)\} = {\rm min}\left\{ \frac{2 e^{-2 p^2}}{\pi \left(1 + e^{-2 |\beta|^2} \right)} \left[e^{-2 |\beta|^2} + \cos(4 \beta p) \right] \right\}.
\end{equation}
The minimum of this function can be solved numerically and for $\beta=2$ we find $\W+ \approx -0.476$.

\subsection{Cat State Generation}

Though there are a number of proposed mechanisms to generate cat states in harmonic oscillator systems, here we focus on two protocols that have been recently demonstrated in superconducting microwave cavities using reservoir engineering \cite{leghtas_2015_si} and Kerr nonlinearities \cite{grimm_2020_si}, as these two methods are the most relevant to our protocol.

\subsubsection{Reservoir Engineering}
\label{decat}

As was originally shown by Leghtas \etal~using superconducting circuits \cite{leghtas_2015_si}, one method to generate cat states is to engineer the dissipation of the system such that it evolves according to the master equation \cite{mirrahimi_2014_si}
\begin{equation}
\label{MErecat}
\dot{\rho} = -\frac{i}{\hbar} [H_{\rm sq},\rho] + \Gamma_2 L [b^2] \rho,
\end{equation}
where
\begin{equation}
\label{Hrecat}
\frac{H_{\rm sq}}{\hbar} = \epsts b^2 + \epst b^\dag{}^2,
\end{equation}
is the squeezing Hamiltonian for the harmonic oscillator mode with annihilation operator $b$ and squeezing amplitude $\epst$. Here, we have also introduced the Lindblad superoperator $L[o] \rho = o \rho o^\dag - \frac{1}{2} o^\dag o \rho - \frac{1}{2} \rho o^\dag o$ (for an arbitrary operator $o$ and density matrix $\rho$) that characterizes the engineered two-quanta losses from the system at a rate $\Gamma_2$. One can also write Eq.~\eqref{MErecat} in the equivalent form \cite{asjad_2014_si}
\begin{equation}
\label{MErecat2}
\dot{\rho} = \Gamma_2 L [b^2 - \beta^2] \rho,
\end{equation}
where $\beta^2 = -2i \epst / \Gamma_2$. The steady state solution $\ket{\psi}_s$ of this master equation is found according to \cite{tan_2013a_si}
\begin{equation}
\label{psiss}
\left( b^2 - \beta^2 \right) \ket{\psi}_s = 0 \Rightarrow b^2 \ket{\psi}_s = \beta^2 \ket{\psi}_s,
\end{equation}
which is the eigenstate of the $b^2$ operator with eigenvalue $\beta^2$ (\eg~a cat state of size $\beta$). Therefore, if one initially prepares the system into an even (odd) Fock state, the two-quanta loss and squeezing processes will evolve this initial state into an even (odd) cat state of size \cite{mirrahimi_2014_si}
\begin{equation}
\label{betaDE}
\beta = \sqrt{\frac{2 |\epst|}{\Gamma_2}},
\end{equation}
on a timescale set by $\tau \sim 1 / \Gamma_2$ \cite{chamberland_2022_si}. 

\subsubsection{Kerr Cats}

An alternate method to generate a cat state in a harmonic oscillator is using a Hamiltonian of the form \cite{grimm_2020_si}
\begin{equation}
\label{HKerr}
\frac{H_{\rm Kerr}}{\hbar} = \epst b^\dag{}^2 + \epsts b^2 - K b^\dag{}^2 b^2 = -K \left(b^\dag{}^2 - \frac{\epsts}{K} \right) \left(b^2 - \frac{\epst}{K} \right) + \frac{|\epst|^2}{K}.
\end{equation}
This becomes obvious when written in the factored form on the right hand side of the equation, as one can easily see that the even and odd cat states of size 
\begin{equation}
\label{betaKC}
\beta_{\rm kc} = \sqrt{\frac{|\epst|}{K}},
\end{equation}
are ground states of this Hamiltonian.

\subsection{Decoherence}

Up to this point, we have discussed pure cat states that exist in isolation from their environment. However, any realistic system will suffer loss and decoherence due to interactions with its surrounding bath. These processes can be described as single-quanta losses to and incoherent excitations from this bath at a rate $\Gamma_1$. Mathematically, these effects can be modelled via the master equation
\begin{equation}
\label{MEcatdec}
\dot{\rho} = \Gamma_1 \left(\nth + 1 \right) L[b] \rho + \Gamma_1 \nth L[b^\dag] \rho,
\end{equation}
where the first (second) term accounts for losses to (excitations from) the bath of average thermal occupancy $\nth$. If one were to seed the above master equation with an initial pure even or odd cat state [\ie~the density matrices associated with the states given in Eq.~\eqref{catstate}], the Wigner function of the resulting state will evolve in time according to \cite{kim_1992_si}
\begin{equation}
\label{Wigcatdec}
W_\pm(x,p,t) = \frac{2 e^{-2 ( x^2 + p^2) / \xi(t) }}{ \pi \left(1 \pm e^{-2 |\beta|^2} \right) \xi(t)} \left\{ \cosh \left( \frac{4 e^{-\Gamma_1 t /2} \beta x}{\xi(t)} \right) \exp \left( \frac{-2 |\beta|^2 e^{-\Gamma_1 t}}{\xi(t)} \right) \pm \cos \left( \frac{4 e^{-\Gamma_1 t/2} \beta p}{\xi(t)} \right) \exp \left( -2 |\beta|^2 \left[1 - \frac{e^{-\Gamma_1 t}}{\xi(t) } \right] \right) \right\},
\end{equation}
where again the plus (minus) sign corresponds to an even (odd) cat state. Here, we have defined $\xi(t) = 2 \nth \left(1 - e^{-\Gamma_1 t} \right) + 1$, such that $\xi(t) = 1$ for a bath in its ground state, while exponentially approaching twice the bath occupancy on a timescale set by $t \sim 1 / \Gamma_1$ for a bath with $\nth \gg 1$. By inputting $t=0$ into Eq.~\eqref{Wigcatdec}, while also taking $\xi(t) = 1$, one immediately recovers the expected Wigner function of a pure cat state given by Eq.~\eqref{Wigcat}. Meanwhile, the steady state state solution is simply a Gaussian thermal state [mathematically, this can be seen by taking $t \rightarrow \infty$ in Eq.~\eqref{Wigcatdec}], indicating that the cat state has lost its initial energy and coherence to its environment.

We conclude this section by noting that the single-quanta processes given in Eq.~\eqref{MEcatdec} will act to hinder any attempts to produce a coherent cat state by introducing loss and decoherence into the system. For instance, in order to generate high-fidelity cat states using the dissipatively driven method discussed in Section \ref{decat}, one must ensure that the engineered two-quanta loss of the system overwhelms these linear processes (\ie~$\Gamma_2 \gg \Gamma_1$). However, it is important to note that even in this situation, where high fidelity cat states can be generated on short timescales $t \sim 1 / \Gamma_2$, in the steady state the interference fringes of the cat state will eventually disappear. One can imagine this decoherence process as a random parity flipping of the cat state due to single-quanta processes, which occur at the rate $\Gamdec = 2 |\beta|^2 \Gamma_1 \left( 2 \nth + 1 \right)$ \cite{kim_1992_si}. On timescales much larger than $1 / \Gamdec$, this parity flipping will result in equal probability of even and odd cat states, whose interference fringes average together and cancel each other out. The remaining state will then be a classical statistical mixture of two coherent states 180 degrees out of phase with each other described by the Wigner function \cite{gerry_1997_si}
\begin{equation}
\label{Wigcm}
\overline{W}(x,p) = \frac{2}{\pi} e^{-2 ( x^2 + p^2) } \cosh(4 \beta x) e^{-2 | \beta|^2},
\end{equation}
which is always positive. This representation is simply the Wigner function of the cat states given in Eq.~\eqref{Wigcat} where the coherence associated with the cosine term has vanished, such that both even and odd cat states approach the same statistical mixture state.

\subsection{Approximate Analytical Model for Cat State Generation via Reservoir Engineering}

\begin{figure}[h!]
\centerline{\includegraphics[width=4in]{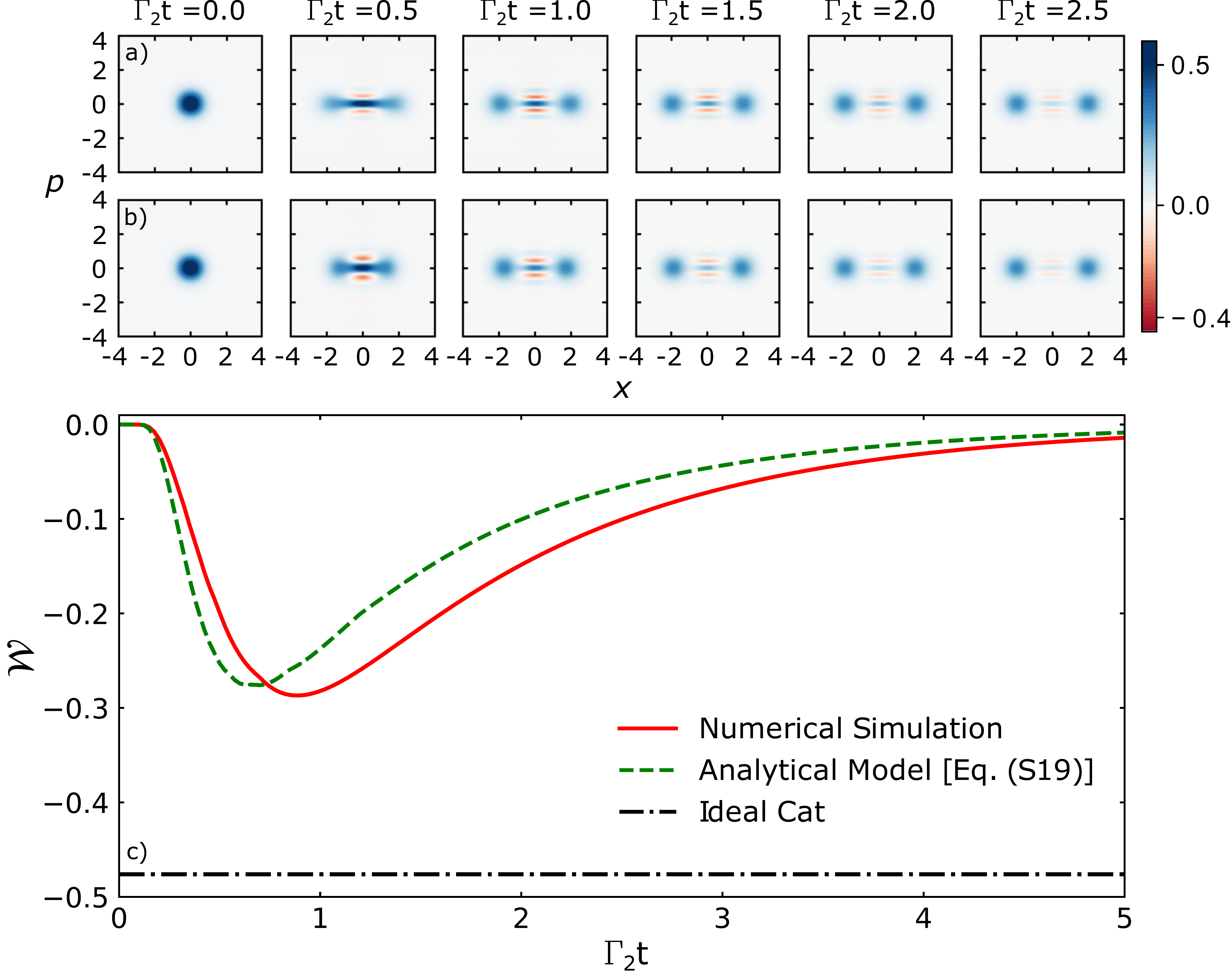}}
\caption{{\label{figs2}} Evolution of a mechanical resonator's Wigner function from its ground state to an even cat state calculated by a) numerically solving the master equation given by Eq.~\eqref{MErecatfull} and b) using the approximate analytical form in Eq.~\eqref{Wigreapp} with Eq.~\eqref{betat} for $\beta_t(t)$. Here, we have taken $\Gamma_1 = 0.1$ and $\Gamma_2 = 1$, with $\epsilon_2$ chosen such that the final cat state reaches a size of $\beta = 2$. In c), we show the Wigner negativity of the mechanical state vs time, with the solid red and green dashed lines corresponding to the numerical simulation given in a) and the analytical approximation given in b), respectively. The black dashed-dotted line corresponds the the Wigner negativity of an ideal even cat state of size $\beta = 2$ ($\W+ \approx -0.476$).}
\end{figure}

We are now poised to develop an approximate analytical theory to model the generation of cat states using reservoir engineering. Here, we consider a harmonic oscillator that is initially prepared into its ground state before evolving into an even cat state according to the protocol detailed in Sec.~\ref{decat} in the presence of environmental decoherence. We model this process by combining Eqs.~\eqref{MErecat} and \eqref{MEcatdec} with $\nth$ set to zero, resulting in the master equation
\begin{equation}
\label{MErecatfull}
\dot{\rho} = - \frac{i}{\hbar} [H_{\rm sq},\rho] + \Gamma_2 L [b^2] \rho + \Gamma_1 L[b] \rho.
\end{equation}
An approximate solution for the time-varying Wigner function that results from this master equation can be formulated by modifying Eq.~\eqref{Wigcatdec} in the following ways:
\begin{enumerate}

  \item Set $\xi(t) = 1$ as we are considering the situation of zero bath temperature (\ie~$\nth=0$).
  
  \item Consider times $t \ll 1/\Gamma_1$ such that we can expand the exponentials with $\Gamma_1 t$ in their arguments, while only keeping leading order terms.
  
  \item Allow the size of the cat state to vary in time. Formally, this is done by replacing $\beta$ by $\beta_t(t)$ with the boundary conditions $\beta_t(0) = 0$ (we initially start in the ground state) and $\beta_t(t \rightarrow \infty) = \beta$. This second condition is justified by the fact that for long times $t \gg \Gamma_2$, the oscillator will evolve into a cat state (or statistical mixture of coherent states) of size $\beta \approx \sqrt{2 |\epst| / \Gamma_2}$ -- see Eq.~\eqref{betaDE}. 
  
\end{enumerate}
Applying these conditions to Eq.~\eqref{Wigcatdec}, we find the approximate time-dependent Wigner function that describes the cat state evolution under our reservoir engineering procedure as
\begin{equation}
\label{Wigreapp}
W_\pm(x,p,t) \approx \frac{2 e^{-2 ( x^2 + p^2)}}{ \pi \left(1 \pm e^{-2 |\beta_t(t)|^2} \right)} \left\{ \cosh \left[ 4 \beta_t(t) x \right] e^{-2 |\beta_t(t)|^2} \pm \cos \left[ 4 \beta_t(t) p \right] e^{-2 |\beta_t(t)|^2 \Gamma_1 t } \right\}.
\end{equation}
We note that by inputting a time-dependent $\beta_t(t)$ into Eq.~\eqref{Wigcatdec} and using this as an ansatz to a differential equation for the Wigner function generated by Eq.~\eqref{MErecatfull}, one may be able to come up with an exact analytical solution for $\beta_t(t)$, and therefore $W_\pm(x,p,t)$, for this system. However, a simple choice would be to model the evolution of $\beta(t)$ in time using the exponential solution
\begin{equation}
\label{betat}
\beta_t(t) = \beta \left( 1 - e^{- 2 \Gamma_2 t} \right).
\end{equation}
Such a choice of $\beta_t(t)$ maps the oscillator's ground state into an even cat state and qualitatively matches the exact solution found using numerical simulation of Eq.~\eqref{MErecatfull}, as can be seen in Fig.~\ref{figs2}.

Finally, we can use the approximate Wigner function given by Eq.~\eqref{Wigreapp} to better understand how its minimal value evolves in time when varying the strengths of $\Gamma_1$ and $\Gamma_2$. To do this, we assume that the Wigner negativity will be minimized at time
\begin{equation}
\label{tmin}
t_{\rm min} = \frac{C}{\Gamma_2} \left( \frac{\Gamma_1}{\Gamma_2} \right)^k.
\end{equation}
Here, we have taken inspiration from the inset of Fig.~3 in the main text such that $t_{\rm min}$ is on the order of $1 / \Gamma_2$, with a dimensionless prefactor $C$, while allowing for a weak power law dependence on $\Gamma_1 / \Gamma_2$ with exponent $k$. We continue by assuming that we are generating a reasonably large cat state ($\beta \geq 2$) and that $\beta_t(t_{\rm min}) \approx \beta$. This second assumption is justified when considering the approximate form of $\beta_t(t)$ given in Eq.~\eqref{betat}, as $\beta_t(t_{\rm min} \sim 1 / \Gamma_2) \approx \beta (1 - e^{-2}) \approx \beta$ in this case. With these two assumptions, the minimal Wigner negativity will exist in the tails of the two classical Gaussian lobes centered at $x = \pm |\beta|$ defined by the first term under the brackets in Eq.~\eqref{Wigreapp}. In this situation, we can therefore neglect this term and express the minimal Wigner negativity vs time as
\begin{equation}
\label{Wnegmint}
\mathcal{W}_{\rm min} \equiv {\rm min} \left\{ W_\pm(x,p,t_{\rm min}) \right\} \approx \mathcal{W}_{\pm} \exp \left[ -2 C |\beta|^2 \left( \frac{\Gamma_1}{\Gamma_2} \right)^{k+1} \right],
\end{equation}
where we have used the definition for $\mathcal{W}_{\pm}$ given by Eqs.~\eqref{Wignego} and \eqref{Wignege}. Thus, with the simple (and often valid) assumptions used in this section, we have shown that the minimal Wigner negativity of the generated cat state decays exponentially in proportion to $\left( \Gamma_1 / \Gamma_2 \right)^{k+1}$. As can be seen in Fig.~3 of the main text, this exponential behaviour is valid over multiple orders of magnitude in $\Gamma_1 / \Gamma_2$. Therefore, provided one knows the parameters $C$ and $k$, one can use Eq.~\eqref{Wnegmint} to determine the minimal Wigner negativity of the cat state of size $\beta$ generated using this protocol. Note that when two-quanta losses dominate their linear counterpart (\ie~$\Gamma_2 \gg \Gamma_1$), $\mathcal{W}_\pm(t_{\rm min}) \approx \mathcal{W}_\pm$ and a near ideal cat state is generated with high fidelity.

\newpage

\newpage

\section{Reduced Model for Optomechanically Engineered Mechanical Cat States}
\label{redmod}

Here, we present the detailed calculations associated with determining the reduced model for the mechanical resonator in the doubly-pumped optomechanical cavity studied in the main text. Starting with the full optomechanical Hamiltonian of the system, we perform a Schrieffer-Wolff transformation, followed by a procedure to adiabatically eliminate the cavity mode, thus arriving at a reduced master equation for our mechanical resonator. Upon investigation of this final result, we deduce that such a system can be used to dissipatively engineer macrosopic mechanical cat states, as described in the main text.

\subsection{Optomechanical Hamiltonian}

We begin with the general Hamiltonian for an optomechanical system that is pumped by two separate tones at different frequencies:

\begin{equation}
\label{fullham1}
\frac{H}{\hbar} = \omc a^\dag a + \omm b^\dag b + g_0 a^\dag a (b + b^\dag) + \left( \epsp a^\dag e^{-i \omp t} + \epsps a e^{i \omp t} \right) + \left( \epsd a^\dag e^{-i \omd t} + \epsds a e^{i \omd t} \right).
\end{equation}
From left to right, each of the terms above describe the following:

\begin{enumerate}

  \item The electromagnetic cavity with resonant frequency $\omc$ and annihilation/creation operators $a$/$a^\dag$.
  
  \item The mechanical resonator with resonant frequency $\omm$ and annihilation/creation operators $b$/$b^\dag$.
  
  \item The optomechanical interaction with single-photon, single-phonon (vacuum) optomechanical coupling rate $g_0$.
  
  \item The strong pump applied to the cavity with frequency $\omp$ and amplitude $\epsp$.
  
  \item The weak drive applied to the cavity with frequency $\omd$ and amplitude $\epsd$.
  
\end{enumerate}

Next we simultaneously transform our system into the frame rotating at the cavity pump frequency and displace the cavity by its steady state amplitude $\alpha = \epsp / (\Delta + i \kappa /2)$ (this amplitude only accounts for cavity photon population due to the pump tone -- a small amount of additional photons will be added by the weak cavity drive as well). Here, $\Delta = \omp - \omc$ is the detuning of the pump frequency from cavity resonance and $\kappa$ is the loss rate of the cavity \cite{tan_2013a_si, asjad_2014_si, machado_2016_si}. Note that this transform is equivalent to applying the unitary $U = e^{i \omp a^\dag a t} e^{\alpha^* a - \alpha a^\dag}$ to the Hamiltonian in Eq.~\eqref{fullham1}. The resulting Hamiltonian in the transformed frame is then
\begin{align}
\label{fullham2}
&\frac{H}{\hbar} = -\Delta a^\dag a + \omm b^\dag b + \epsd a^\dag e^{i \Delp t} + \epsds a e^{-i \Delp t} + g_0 (\alpha^* a + \alpha a^\dag) (b + b^\dag) + g_0 a^\dag a (b + b^\dag),
\end{align}
where $\Delp = \omp - \omd$ is the detuning between the cavity's pump and drive frequencies. In performing this transformation, we have effectively absorbed the effects of the pump into $\alpha$, allowing us to break the optomechanical interaction into its linear (second-last term) and nonlinear (last term) parts. In this displaced frame the operators $a$ and $a^\dag$ now represent the ladder operators of the displaced cavity mode, which will in general be in a thermal state. We further note that we have implicitly accounted for the {\red displacement in} equilibrium position of the mechanical resonator, along with the resulting static shift {\red of the} cavity frequency {\red by $\delta \omc = 2 g_0^2 |\alpha|^2 / \omm$}, to omit the term $g_0 |\alpha|^2 (b + b^\dag)$ in Eq.~\eqref{fullham2} associated with the average radiation pressure force applied to the mechanical resonator \cite{aspelmeyer_2014_si}. {\red In practice, one can account for this effect by modifying the detuning of the pump by an amount $\delta \omc$.} We stress that as we have only performed unitary operations and inconsequential static shifts of the resonator's equilibrium position and the cavity's frequency, the Hamiltonian in Eq.~\eqref{fullham2} represents a general description of our optomechanical system. This allows for exact numerical simulations on a Hilbert space considerably smaller than the one that would be required to contain the pump degrees of freedom, resulting in a significant increase in computation efficiency \cite{chamberland_2022_si}.

\subsection{Schrieffer-Wolff Transformation}

In general, a Schrieffer-Wolff transformation is applied to the Hamiltonian of a system in order to diagonalize it to arbitrary order in some small parameter \cite{luttinger_1955_si, schrieffer_1966_si}. As a concrete example, imagine a Hamiltonian of the form 
\begin{equation}
\label{exHam}
H = H_0 + V,
\end{equation}
where $H_0$ is the diagonalized portion of the Hamiltonian, while $V$ represents a small off-diagonal perturbation. To perform a Schrieffer-Wolff transformation on this Hamiltonian we apply a unitary transformation with generator $S$ (which will be small on the same order as $V$), resulting in
\begin{equation}
\label{SWtrans}
H' = e^S H e^{-S} \approx H_0 + V + [S,H_0] + [S,V] + \frac{1}{2} [S,[S,H_0]],
\end{equation}
where we have only kept terms to second order in $V$ and $S$. By choosing $S$ such that $[S,H_0] = -V$, the first-order non-diagonal term $V$ is removed, resulting in the Hamiltonian
\begin{equation}
\label{SWtrans2}
H' \approx H_0 + \frac{1}{2} [S,V],
\end{equation}
which is diagonalized to second order in $V$ and $S$.

We can apply this type of transformation to the optomechanical Hamiltonian in Eq.~\eqref{fullham2} to investigate the nonlinear effects of the optomechanical interaction to first order in the small parameter $g_0/\omm$. To do this, we take 
\begin{equation}
\label{H0}
H_0 = - \hbar \Delta a^\dag a + \hbar \omm b^\dag b + \hbar \epsd a^\dag e^{i \Delp t} + \hbar \epsds a e^{-i \Delp t} + \hbar g_0 (\alpha^* a + \alpha a^\dag) (b + b^\dag),
\end{equation} 
to be the Hamiltonian excluding the nonlinear optomechanical interaction
\begin{equation}
\label{V}
V = \hbar g_0 a^\dag a (b + b^\dag),
\end{equation}
which we here treat as a perturbation with a generator of the form 
\begin{align}
\label{SWSop}
S = \frac{g_0}{\omm} a^\dag a (b^\dag - b).
\end{align}
We note that while this generator does not produce a true Schrieffer-Wolff transformation in the sense that it does not completely diagonalize our system, it can be used to diagonalize the optomechanical Hamiltonian in Eq.~\eqref{fullham1} in the absence of the drive and damping terms \cite{nunnenkamp_2011_si}.

Evaluating each of the commutators in Eq.~\eqref{SWtrans} we find \cite{machado_2016_si}
\begin{align}
\label{firstcomm}
&[S,\hbar \Delta a^\dag a] = 0, \\
\label{secondcomm}
&[S,\hbar \omm b^\dag b] = - \hbar g_0 a^\dag a \left( b + b^\dag \right) = -V, \\
\label{thirdcomm}
&[S,\hbar \epsd a^\dag e^{i \Delp t} + \hbar \epsds a e^{-i \Delp t}] = \frac{\hbar g_0}{\omm} \left\{ \epsd a^\dag e^{i \Delp t} - \epsds a e^{-i \Delp t} \right\} (b^\dag - b), \\
\label{fourthcomm}
&[S,\hbar g_0 (\alpha^* a + \alpha a^\dag) (b + b^\dag)] = \frac{\hbar g_0^2}{\omm} \left\{ \alpha^* \left(b^2 - b^\dag{}^2 -2 a^\dag a -1 \right) a + \alpha a^\dag \left(b^\dag{}^2 - b^2 -2 a^\dag a - 1 \right)  \right\}, \\
\label{fifthcomm}
&[S,V] = - \frac{2 \hbar g_0^2}{\omm} \left( a^\dag a \right)^2, \\
\label{lastcomm}
&\frac{1}{2} [S,[S,H_0]] \approx - \frac{1}{2} [S,V] = \frac{\hbar g_0^2}{\omm} \left(a^\dag a \right)^2.
\end{align}
Note that in the double commutator found in Eq.~\eqref{lastcomm}, we have only kept terms up to $g_0^2/\omm$ (\ie~we have neglected terms of order $(g_0/\omm)^2$ or higher). Sure enough, the commutator in Eq.~\eqref{secondcomm} produces a term that exactly cancels the nonlinear part of the optomechanical interaction. Therefore, when we input each of the commutation relations of Eqs.~\eqref{firstcomm}--\eqref{lastcomm} into Eq.~\eqref{SWtrans}, we arrive at the nonlinear optomechanical Hamiltonian to first order in $g_0 / \omm$ as 
\begin{equation}
\begin{split}
\label{SWHam}
\frac{H'}{\hbar} &= -\Delta a^\dag a + \omm b^\dag b + \left( \gos a + \go a^\dag \right) \left(b + b^\dag \right) +  \left( \gts a - \gt a^\dag \right) \left(b^2 - b^\dag{}^2 \right)  + \epsd a^\dag e^{i \Delp t} + \epsds a e^{-i \Delp t} \\
&+ \frac{g_0}{\omm} \left\{ \epsd a^\dag e^{i \Delp t} - \epsds a e^{-i \Delp t} \right\} (b^\dag - b) - \frac{g_0^2}{\omm} \left\{ \left(a^\dag a \right)^2 + \alpha^* \left(2 a^\dag a + 1 \right) a + \alpha a^\dag \left( 2 a^\dag a + 1 \right)  \right\},
\end{split}
\end{equation}
where we have introduced the first- and second-order cavity-enhanced optomechanical coupling rates $\go = g_0 \alpha$ and $\gt = g_0^2 \alpha / \omm = g_0 \go / \omm$. The first line of this Hamiltonian contains the first- and second-order optomechanical processes whereby either one or two phonons in the mechanical resonator are annihilated (created) to upconvert (downconvert) a pump photon into a cavity photon. These are the terms that will be relevant to our cat state generation protocol. Meanwhile, the second line consists of higher-order corrections to the cavity's response, which can be neglected in the regime where $g_0 \ll \omm$ and $\braket{a} \ll 1$ (in the next section we shall see that this second condition is also necessary to adiabatically eliminate the cavity mode). We can therefore write the master equation of our system as \cite{tan_2013a_si, asjad_2014_si} 
\begin{equation}
\label{SWME}
\dot{\rho} = \liou \rho = \left( \liou_a + \liou_b + \liou_i \right) \rho,
\end{equation}
where we have introduced a Liouvillian superoperator $\liou = \liou_a + \liou_b + \liou_i$ to account for coupling of our system with its electromagnetic and mechanical baths ($\liou_a$ and $\liou_b$, respectively), as well as the coherent optomechanical interaction ($\liou_i$), which are in general defined as
\begin{align}
\label{Lioua}
&\liou_a \rho = \kappa (\na + 1) L[a] \rho + \kappa \na L[a^\dag] \rho, \\
\label{Lioub}
&\liou_b \rho = \Gamma (\nb + 1) L[b] \rho + \Gamma \nb L[b^\dag] \rho, \\
\label{Lioui}
&\liou_i \rho = - \frac{i}{\hbar} [H'',\rho].
\end{align}
Here, $\na$ ($\nb$) is the average thermal photon (phonon) occupation of the electromagnetic cavity (mechanical resonator) and $H'' = H_1 + H_2 + H_3$ is the Hamiltonian describing the coherent evolution of the system, which is broken into three parts corresponding to the first- and second-order optomechanical interactions and the cavity drive, respectively. We also remind the reader that we have chosen to define our Lindblad superoperator as $L[o] \rho = o \rho o^\dag - \frac{1}{2} o^\dag o \rho - \frac{1}{2} \rho o^\dag o$. By transforming into the frame that rotates at the cavity detuning and mechanical frequency (\ie~by applying the unitary $U = e^{-i \Delta a^\dag a t} e^{i \omm b^\dag b t}$ to $H''$), we can write each of these terms as
\begin{align}
\label{Ham1}
&H_1 = \hbar \left( \gos a e^{i \Delta t} + \go a^\dag e^{-i \Delta t} \right) \left(b e^{-i \omm t} + b^\dag e^{i \omm t} \right), \\
\label{Ham2}
&H_2 = \hbar \left( \gts a e^{i \Delta t} - \gt a^\dag e^{-i \Delta t} \right) \left(b^2 e^{-i 2 \omm t} - b^\dag{}^2 e^{i 2 \omm t} \right), \\
\label{Ham3}
&H_3 = \hbar \left( \epsd a^\dag e^{i \Delc t} + \epsds a e^{-i \Delc t} \right),
\end{align}
where $\Delc = \omc - \omd$ is now the detuning of the cavity's drive from its resonance frequency.

\subsection{Adiabatic Elimination of the Cavity Mode}

With our Schrieffer-Wolff transformation complete, we now move on to adiabatically eliminate the fast cavity mode from the system by tracing over the optomechanical interaction and leaving behind an effective description of the mechanical system. Here, we use the Nakajima-Zwanzig formalism \cite{nakajima_1958_si, zwanzig_1960_si, zwanzig_1964_si}, as was previously used by Wilson-Rae \etal~\cite{wilsonrae_2008_si} and L\"orch \cite{lorch_2015_si} to investigate linear quantum effects such as optomechanical sideband cooling and phonon lasing.

To perform this calculation, we first define the projection operator 
\begin{equation}
\label{Pop}
P \rho = \lim_{t \rightarrow \infty} e^{\liou_a t} \rho = \bar{\rho}_a \otimes \rho_b,
\end{equation}
which projects the system into the slow subspace comprised of the cavity in a thermal state governed by the steady-state density matrix
\begin{equation}
\label{rhoa}
\bar{\rho}_a = \frac{1}{\na + 1} \sum_{n=0}^\infty \left[\frac{\na}{\na + 1} \right]^n \ket{n} \bra{n},
\end{equation}
and the density matrix of the mechanical system $\rho_b = \Tr \{ \rho \}$. Here, $\Tr\{\}$ is defined as the trace over the cavity degrees of freedom. Note that we have chosen the steady state of the cavity to be thermal, though we can easily choose this steady state to be the ground state by setting $\na = 0$. 

We also define the orthogonal projection operator $Q = I - P$, where $I$ is the identity operation on the entire system, such that $Q$ projects the system into the fast subspace spanned by the cavity's excited states. These two projection operators therefore have the following properties:
\begin{align}
\label{PQI}
&P + Q = I, \\
&PQ = QP = 0, \\
&P^m = P, \\
\label{Qn}
&Q^m = Q,
\end{align}
where $m$ is any integer. We can then insert the identity given in Eq.~\eqref{PQI} into Eq.~\eqref{SWME} to obtain master equations for both the slow and fast subspaces as
\begin{align}
\label{PME}
&P \dot{\rho} = P \liou (P + Q) \rho, \\
\label{QME}
&Q \dot{\rho} = Q \liou (P + Q) \rho.
\end{align}
To proceed, we formally integrate Eq.~\eqref{QME} for the fast subspace resulting in
\begin{equation}
\label{Qint}
Q \rho(t) = G(t,t_0) Q \rho(t_0) + \int_{t_0}^t dt' G(t,t') Q \liou(t') P \rho(t'),
\end{equation}
where $G(t,t')$ is some yet-to-be-determined propagator \cite{lorch_2015_si}. If we assume that the system is initially in the $P$ subspace (or at least that $\displaystyle \lim_{t \rightarrow \infty} G(t,t_0) Q = 0$ -- \ie~if we wait long enough, any initial part of the state that was in the $Q$ subspace dies off on the relevant timescale of the system), then we can ignore the first term in Eq.~\eqref{Qint} and insert it into Eq.~\eqref{PME} to obtain
\begin{equation}
\label{Pint}
P \dot{\rho} = P \liou P \rho + P \liou \int_{t_0}^t dt' G(t,t') Q \liou(t') P \rho(t').
\end{equation}

In order to solve Eq.~\eqref{Pint}, we now invoke our adiabatic approximation. First, if $Q \liou_a Q$ has all real, negative eigenvalues (this will be true for any physical system by construction) and these eigenvalues are larger in magnitude than $||Q \liou_b Q ||$ and $||Q \liou_i Q ||${\red , then} the dynamics of $\liou_a$ {\red will} set the fast timescale {\red of the system. This statement is equivalent to ensuring that the loss rate of the electromagnetic cavity $\kappa$ exceeds the coupling rate of the mechanical resonator to its own thermal bath, as well as its optomechanical coupling to the cavity. Mathematically, these conditions amount to the limits $g_1 \ll \omm$ and $2 g_2, \Gamma (\nth + 1) \ll \kappa$ (this will be made clear below).} {\red In this situation,} we can approximate the propagator introduced in Eq.~\eqref{Qint} as
\begin{equation}
\label{Gapprox}
G(t,t') Q \approx Q e^{\liou_a (t-t')} Q. 
\end{equation}
We note that in some references the propagator is equivalently written as $G(t,t') Q \approx e^{ Q \liou_a (t-t')} Q$ \cite{lorch_2015_si}, which is a result of the fact that $e^{ Q \liou_a (t-t') Q} = e^{ Q \liou_a (t-t')} Q = Q e^{\liou_a (t-t') Q} = Q e^{\liou_a (t-t')} Q$ by definition of the exponential operator and the projection operator property given in Eq.~\eqref{Qn}.

Along with this approximation for the propagator, we make the assumption that $P \rho(t') \approx P \rho(t)$. That is to say the operation of $P$ on the density matrix is independent of time, which is justified by the fact that $P$ defines the subspace that varies in time slowly. Using these two approximations, along with a change of variables $\tau = t-t'$, we can rewrite the master equation for the slow subspace as
\begin{equation}
\label{Pintapp}
P \dot{\rho} (t) = P \liou(t) P \rho(t) + P \liou(t) Q \int_0^\infty d\tau e^{\liou_a \tau} Q \liou(t - \tau) P \rho(t).
\end{equation}
Here, we have explicitly written out the time dependence of the Liouvillians and density matrices to avoid confusion, as well as translated the integral limits to $t_0 = 0$ and $t \rightarrow \infty$ in the spirit of the adiabatic approximation.

We now make a number of observations to help simplify the system \cite{lorch_2015_si}:

\begin{enumerate}

  \item $\liou_a P \rho = 0$: This is because $\liou_a \bar{\rho}_a = 0$ due to the fact that $\bar{\rho}_a$ describes the steady state of the cavity.
  
  \item $P \liou_a = 0$: Along with the previous point, this is required in order to preserve the trace over the cavity.
  
  \item $Q \liou_b P = P \liou_b Q = 0$: The projections don't act on the (slow) mechanical subspace so all three of these operators commute and $PQ=0$.
  
  \item $P \liou_i P=0$: $\liou_i$ describes an interaction that takes the system between the two subspaces, not from the slow subspace back onto itself.
  
\end{enumerate}
Inputting each of these conditions into Eq.~\eqref{Pintapp}, while using the fact that $Q = I - P$, we obtain a simplified expression for the master equation of the slow subspace as
\begin{equation}
\label{Pintsimp}
P \dot{\rho} (t) = P \liou_b(t) P \rho(t) + P \liou_i(t) \int_0^\infty d\tau e^{\liou_a \tau} \liou_i(t - \tau) P \rho(t).
\end{equation}
Here, the first term simply accounts for the intrinsic losses of the mechanical system. Therefore, we are mainly interested in the second term, which can be evaluated by inputting Eq.~\eqref{Pop} to arrive at
\begin{equation}
P \liou_i(t) \int_0^\infty d\tau e^{\liou_a \tau} \liou_i(t - \tau) P \rho(t) = \Tr \left\{ \liou_i(t) \int_0^\infty d \tau e^{\liou_a \tau} \liou_i (t-\tau) \bar{\rho}_a \otimes \rho_b \right\} \otimes \bar{\rho}_a.
\end{equation}
Our calculation then amounts to using Eq.~\eqref{Lioui} to evaluate the trace
\begin{equation}
\label{Mijdef}
\begin{split}
\Tr \left\{ \liou_i(t) \int_0^\infty d \tau e^{\liou_a \tau} \liou_i (t-\tau) \bar{\rho}_a \otimes \rho_b \right\} = - \frac{1}{\hbar^2} \Tr \left\{ \left[H''(t), \int_0^\infty d \tau e^{\liou_a \tau} \left[ H''(t - \tau), \bar{\rho}_a \otimes \rho_b \right] \right] \right\} = \sum_{i,j=1}^3 M_{ij},
\end{split}
\end{equation}
where
\begin{equation}
M_{ij} = - \frac{1}{\hbar^2} \Tr \left\{ \left[H_i(t), \int_0^\infty d \tau e^{\liou_a \tau} \left[ H_j(t - \tau), \bar{\rho}_a \otimes \rho_b \right] \right] \right\}.
\end{equation}
Using this notation, we can then write the adiabatically eliminated master equation for the mechanical resonator as
\begin{equation}
\label{mechme}
\dot{\rho}_b = \liou_b \rho_b + \sum_{i,j=1}^3 M_{ij}.
\end{equation}

By invoking the RWA, and assuming that $\Delta \approx - 2 \omm$ and $\Delc \approx 0$ (these are the conditions that will be relevant for our mechanical cat state generation protocol) we can immediately see that $M_{12} = M_{21} = M_{13} = M_{31} = 0$. This is due to the fact that if we multiply $H_1$ by $H_2$ or $H_3$ or vice versa, we will always be left with rotating terms that will average to zero. Furthermore, $M_{33} = 0$, as $H_3$ does not contain any mechanical operators, such that after the trace is performed we will only be left with scalars that will be rendered zero by the commutators in Eq.~\eqref{Mijdef}. While not immediately obvious, it can also be shown that $M_{32} = 0$. We are then left with computing the nonzero terms $M_{11}$, $M_{22}$, and $M_{23}$. Using the relations
\begin{align}
\label{lioua}
e^{\liou_a \tau} a &= e^{- \kappa \tau / 2} a, \\
\label{liouadag}
e^{\liou_a \tau} a^\dag &= e^{- \kappa \tau / 2} a^\dag,
\end{align}
which follow from the fact that $e^{\liou_a \tau}$ evolves the cavity decay forward in time by an amount $\tau$ according to $\liou_a$, we find that calculating $M_{11}$, $M_{22}$, and $M_{23}$ amounts to solving integrals of the form
\begin{equation}
\int_0^\infty d \tau e^{[-\kappa / 2 \pm i (\Delta \pm \omega)] \tau} = \frac{1}{\kappa / 2 \mp i (\Delta \pm \omega)} = \chi_{\mp}(\pm \omega).
\end{equation}
Here, $\chi_-(\omega) = \chi(\omega)$ is the susceptibility of the electromagnetic cavity and $\chi_+(\omega) = \chi^*(\omega)$ \cite{aspelmeyer_2014_si}. Using these relations, while assuming the steady state of the cavity is its ground state ($\na =0$), we find
\begin{align}
\label{M11}
M_{11} = |\go|^2 \left\{ \kappa |\chi(\omm)|^2 L[b] \rho_b + \kappa |\chi(-\omm)|^2 L[b^\dag] \rho_b - i\left[ (\Delta + \omm) |\chi(\omm)|^2 + (\Delta - \omm) |\chi(-\omm)|^2 \right] [b^\dag b,\rho_b] \right\},
\end{align}
\begin{equation}
\begin{split}
\label{M22}
M_{22} &= |\gt|^2 \{ \kappa |\chi(2 \omm)|^2 L[b^2] \rho_b + \kappa |\chi(-2\omm)|^2 L[b^\dag{}^2] \rho_b + i \left[ ( \Delta + 2 \omm) |\chi(2\omm)|^2 - 3 (\Delta - 2 \omm) |\chi(-2\omm)|^2  \right] [b^\dag b,\rho_b] \\ 
&- i\left[ (\Delta + 2\omm) |\chi(2\omm)|^2 + (\Delta - 2 \omm) |\chi(- 2\omm)|^2 \right] [(b^\dag b)^2,\rho_b] \}.
\end{split}
\end{equation}
Here, we note that $M_{11}$ contains the terms associated with the linear optomechanical damping and spring effects. Meanwhile, the first three terms in $M_{22}$ provide the analogous phenomena in the second-order case, with last term in $M_{22}$ introducing a mechanical Kerr nonlinearity. We therefore define a shifted mechanical frequency $\ommt = \omm + \delta \omega_1 + \delta \omega_2$, where $\delta \omega_1$ and $\delta \omega_2$ account for the first- and second-order optomechanical spring effects according to
\begin{align}
\label{dwm1}
&\delta \omega_1 = |\go|^2 \left[(\Delta + \omm) |\chi(\omm)|^2 + (\Delta - \omm) |\chi(-\omm)|^2 \right], \\
\label{dwm2}
&\delta \omega_2 = - |\gt|^2 \left[ ( \Delta + 2 \omm) |\chi(2\omm)|^2 - 3 (\Delta - 2 \omm) |\chi(-2\omm)|^2 \right].
\end{align}
Finally, for $M_{23}$ we find
\begin{equation}
\label{M23}
M_{23} = \frac{\epsd \gts e^{i(\Delc + \Delta + 2 \omm) t}}{\kappa / 2 + i \Delc} [b^\dag{}^2, \rho_b] - \frac{\epsds \gt e^{-i(\Delc + \Delta + 2 \omm) t}}{\kappa / 2 - i \Delc} [b^2, \rho_b].
\end{equation}
By taking $\Delc = 0$ (as will be relevant for our cat state generation procedure), $M_{23}$ simplifies to
\begin{equation}
\label{M23app}
M_{23} = \frac{2}{\kappa} [\epsd \gts b^\dag{}^2 e^{i(\Delta + 2 \omm)t} - \epsds \gt b^2 e^{-i(\Delta + 2 \omm)t}, \rho_b] = -i[\epst b^\dag{}^2 e^{i(\Delta + 2 \omm)t} + \epsts b^2 e^{-i(\Delta + 2 \omm)t}, \rho_b],
\end{equation}
where $\epst = 2 i \epsd \gts / \kappa$ is the two-phonon drive strength of the adiabatically eliminated mechanical system.

\subsubsection{Sideband-Resolved Regime}

Inputting Eqs.~\eqref{M11}--\eqref{M23app} into Eq.~\eqref{mechme}, while taking $\Delc = 0$ and $\Delta = -2 \ommt \approx - 2 \omm$ (that is, pump red-detuned by twice the shifted mechanical frequency to maximize the efficiency of the cat state generation), we find the adiabatically eliminated master equation for the mechanical mode in the sideband-resolved regime ($\kappa \ll \omm$) as
\begin{align}
\label{MEbSBR}
\dot{\rho}_b = -\frac{i}{\hbar} [H_b, \rho_b] + \Gamma \left(\nb + 1 +\frac{\Gamma_1}{\Gamma} \right) L[b] \rho_b +\Gamma \left( \nb + \frac{\Gamma_1}{9 \Gamma} \right) L[b^\dag] \rho_b + \Gamma_2 L[b^2] \rho_b.
\end{align}
Here, we have transformed into the frame rotating at the shifted mechanical frequency, allowing us to introduce the effective mechanical Hamiltonian as
\begin{align}
\label{HbSBR}
\frac{H_b}{\hbar} = \epst b^\dag{}^2 + \epsts b^2 - K (b^\dag b)^2.
\end{align}
In Eq.~\eqref{MEbSBR}, $\Gamma$ is the intrinsic mechanical decay rate, which is modified by the first- and second-order optomechanical damping rates
\begin{align}
\label{Gam1SBR}
&\Gamma_1 = \frac{|\go|^2 \kappa}{\omm^2}, \\
\label{Gam2SBR}
&\Gamma_2 = \frac{4 |\gt|^2}{\kappa} = \frac{4 g_0^2 |\go|^2}{\kappa \omm^2} = \left( \frac{2 g_0}{\kappa} \right)^2 \Gamma_1.
\end{align}
{\red These two optomechanical coupling rates must be small compared to the cavity loss rate $\kappa$ for our adiabatic elimination procedure to be valid, resulting in the conditions $g_1 \ll \omm$ and $2 g_2 \ll \kappa$ highlighted above.} {\red Furthermore,} the optomechanical spring effects become
\begin{align}
\label{dwm1SBR}
&\delta \omega_1 = -\frac{4 |\go|^2}{3 \omm}, \\
\label{dwm2SBR}
&\delta \omega_2 = -\frac{3 |\gt|^2}{4 \omm} = -\frac{3 g_0^2 |\go|^2}{4 \omm^3} = \left( \frac{3 g_0}{4 \omm} \right)^2 \delta \omega_1,
\end{align}
while the mechanical Kerr nonlinearity is evaluated to be
\begin{equation}
\label{KerrSBR}
K = \frac{|\gt|^2}{4 \omm} = \frac{g_0^2 |\go|^2}{4 \omm^3}.
\end{equation}
We note that in the sideband-resolved regime, we have neglected the incoherent, two-phonon drive term $(|\gt|^2 \kappa / 16 \omm^2) L[b^\dag{}^2] \rho_b$ in Eq.~\eqref{MEbSBR}, as it's smaller than the $L[b^2] \rho_b$ term by a factor of $(\kappa / 8 \omm)^2$.

\begin{figure}[h!]
\centerline{\includegraphics[width=5in]{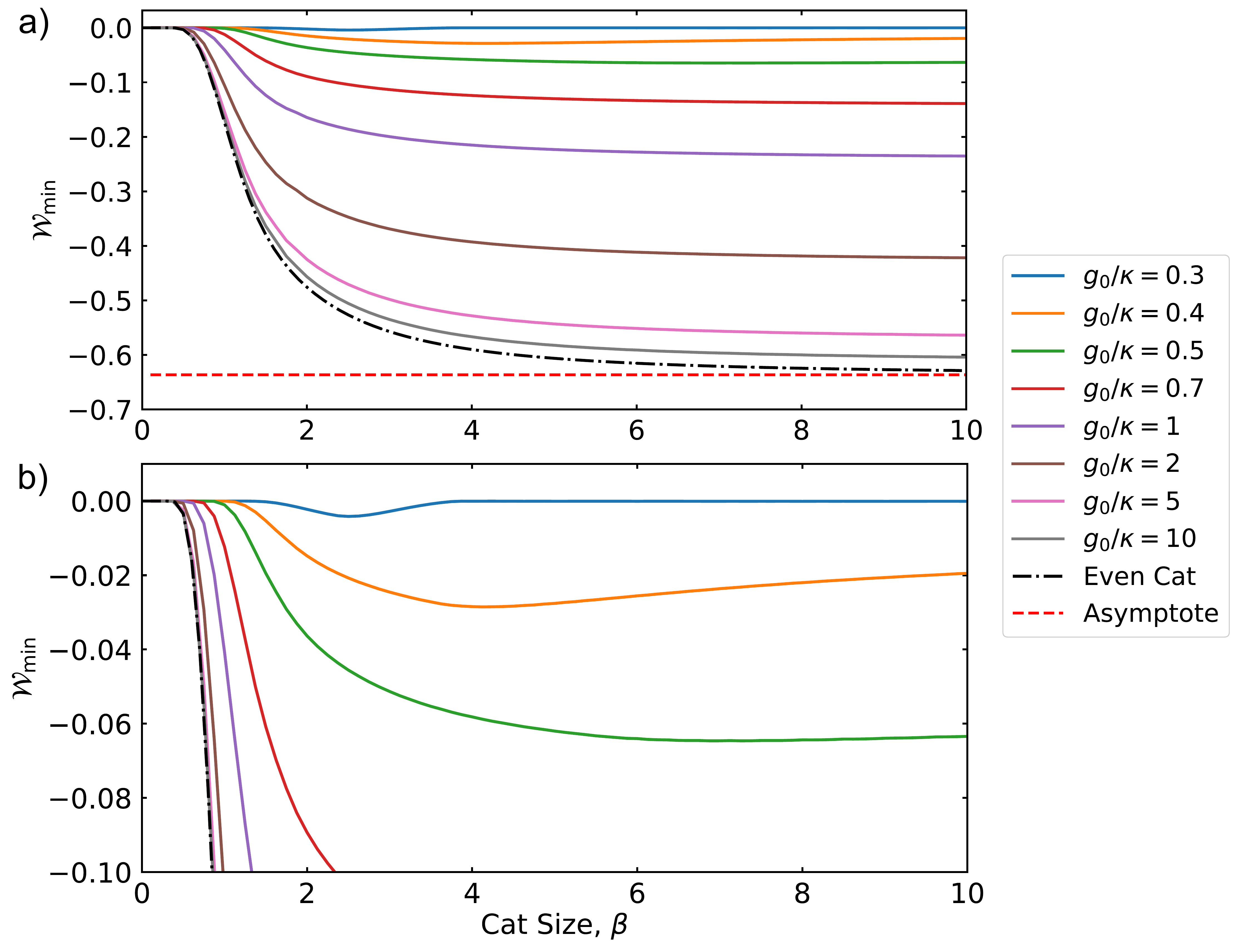}}
\caption{\red {\label{figs3}} a) Plot of minimal Wigner negativity $\mathcal{W}_{\rm min}$ vs cat state size $\beta$ according to the adiabatically eliminated model of Eq.~\eqref{MEbSBR} for varying values of $g_0 / \kappa$ (see legend). In b) we show the same plot zoomed in on the y-axis. All parameters are identical to those used for Fig.~3 of the main text, with $\epsd$ varied to sweep the size of the cat state. The dashed-dotted black line is the minimal Wigner negativity of an ideal even cat state $\mathcal{W}_+$ calculated using Eq.~\eqref{Wignege}, while the red dashed line is the asymptotic value of $- 2 / \pi$ (\ie~the minimal Wigner negativity of an ideal odd cat state $\mathcal{W}_-$).}
\end{figure}

{\red In Fig.~\ref{figs3} we plot the minimal Wigner negativity resulting from the reduced model in Eq.~\eqref{MEbSBR} vs cat state size $\beta$ using the same parameters as Fig.~3 of the main text. Here, we observe that while the minimal value of the Wigner function becomes increasingly negative as $\beta$ gets larger, it is limited to a minimal value whose magnitude shrinks with decreasing $g_0 / \kappa$. This can be seen explicitly in Fig.~\ref{figs3}b, where for $g_0 / \kappa \leq 0.5$ the Wigner negativity begins to increase back towards zero in the plotted range of $\beta \leq 10$. We postulate that this effect would also be seen for larger ratios of $g_0 / \kappa$ for $\beta > 10$, as the decoherence of the cat state scales with $|\beta|^2$. Unfortunately, as the Hilbert space required to simulate our system also grows as $|\beta|^2$, we are unable to consider cat states much larger than $\beta = 10$ to numerically investigate this effect.}

\subsubsection{Non-Sideband-Resolved Regime}

For completeness, we also evaluate the adiabatically eliminated master equation in the non-sideband-resolved regime ($\kappa \gg \omm$), where we find a form similar to Eq.~\eqref{MEbSBR} as
\begin{align}
\label{MEbnonSBR}
\dot{\rho}_b = -\frac{i}{\hbar} [H_b, \rho_b] + \Gamma \left(\nb + 1 +\frac{\Gamma_1}{\Gamma} \right) L[b] \rho_b +\Gamma \left( \nb + \frac{\Gamma_1}{\Gamma} \right) L[b^\dag] \rho_b + \Gamma_2 L[b^2] \rho_b + \Gamma_2 L[b^\dag{}^2] \rho_b,
\end{align}
where $H_b$ takes on the same form as Eq.~\eqref{HbSBR}. The main difference here is that in Eq.~\eqref{MEbnonSBR} the strengths of the incoherent phonon annihilation and creation rates are now comparable so we have included the $L[b^\dag{}^2] \rho_b$ term. Along with these changes, we also have new expressions for the first- and second-order optomechanical dampings, spring effects, and mechanical Kerr nonlinearity as
\begin{align}
\label{Gam1nonSBR}
&\Gamma_1 = \frac{4 |\go|^2}{\kappa}, \\
\label{Gam2nonSBR}
&\Gamma_2 = \frac{4 |\gt|^2}{\kappa} = \frac{4 g_0^2 |\go|^2}{\kappa \omm^2} = \left( \frac{g_0}{\omm} \right)^2 \Gamma_1, \\
\label{dwm1nonSBR}
&\delta \omega_1 = -\frac{16 |\go|^2 \omm}{\kappa^2}, \\
\label{dwm2nonSBR}
&\delta \omega_2 = -\frac{48 \omm |\gt|^2}{\kappa^2} = -\frac{48 g_0^2 |\go|^2}{\kappa^2 \omm} = 3 \left( \frac{g_0}{\omm} \right)^2 \delta \omega_1, \\
\label{KerrnonSBR}
&K = \frac{16 \omm |\gt|^2}{\kappa^2} = \frac{16 g_0^2 |\go|^2}{\kappa^2 \omm}.
\end{align}
However, since we have already assumed $g_0 / \omm \ll 1$ for our analysis, we see $\Gamma_2 \ll \Gamma_1$ (and also $\delta \omega_2 \ll \delta \omega_1$) in this regime, such that nonlinear effects are negligible. Furthermore, in order for the Kerr nonlinearity to be relevant we require $K \gg \Gamma_1$, which amounts to $4 g_0^2 / \kappa \omm \gg 1$. Unfortunately, our assumption that $g_0 \ll \omm \ll \kappa$ in this regime prevents this condition from ever being true. Therefore, in the non-sideband-resolved regime all of the interesting nonlinear effects are washed away, so we only focus on the sideband-resolved regime in this paper.

\newpage

\section{Optomechanical Cat State Engineering}

Comparing Eqs.~\eqref{MEbSBR} and \eqref{HbSBR} to Eqs.~\eqref{MErecat}, \eqref{Hrecat}, and \eqref{HKerr}, we see that our pumping scheme can support cat state generation using both reservoir engineering and a Kerr nonlinearity \cite{gautier_2022_si}. Here, we look at the feasibility of creating mechanical cat states using each of these methods.

\subsection{Reservoir Engineering}

In order to create cat states through reservoir engineering, we require that the nonlinear losses overwhelm their linear counterparts, which amounts to $\Gamma_2 \gg \Gamma_1$, or equivalently, $2 g_0 \gg \kappa$. In this situation, nonlinear processes dominate, allowing for generation of cat states of size
\begin{equation}
\label{betaDEom}
\beta_{\rm de} = \sqrt{\frac{2 |\epst|}{\Gamma_2}} = \sqrt{\frac{|\epsd|}{|\gt|}} {\red \approx \frac{\omega_m}{g_0} \sqrt{\frac{2 |\epsilon_d|}{|\epsilon_p|}}},
\end{equation}
according to Eq.~\eqref{betaDE}. {\red For the approximation on the right-hand side of this equation, we have used $g_2 = |\alpha| g_0^2 / \omega_m$ and $|\alpha| = \sqrt{|\epsilon_p|^2 / (\Delta^2 + \kappa^2 / 4)} \approx |\epsilon_p| / 2 \omega_m$ for the conditions of $\Delta = -2 \omega_m$ and $\omega_m \gg \kappa$ considered in our protocol.} Note that while single-phonon losses associated with $\Gamma_1$ will act to reduce the energy of the system, and therefore its size, Eq.~\eqref{betaDE} is an excellent approximation of the final cat state size in the regime of $\Gamma_2 \gg \Gamma_1$. With the size of the cat state determined, we can also calculate its minimal Wigner negativity for a given ratio $g_0 / \kappa$ by inputting Eq.~\eqref{Gam2SBR} into Eq.~\eqref{Wnegmint} to obtain
\begin{equation}
\label{Wnegmintg0k}
\mathcal{W}_{\rm min} = \mathcal{W}_{\pm} \exp \left[ -2 C |\beta|^2 \left( \frac{\kappa}{2 g_0} \right)^{2k+2} \right],
\end{equation}
where $C$ and $k$ determine the time $t_{\rm min}$ at which the Wigner negativity is minimized according to Eq.~\eqref{tmin}. By fitting the simulation shown in Fig.~3 of the main text, we determine $C \approx 1/3$ and $k \approx -1/4$ for the $\beta = 2$ cat state studied in this work, such that its minimal Wigner negativity exhibits exponential decay proportional to {\red $|\beta|^2 \left( \kappa / 2 g_0 \right)^{3/2}$.}

\subsection{Kerr Nonlinearity}

By comparing the adiabatically eliminated Hamiltonian in Eq.~\eqref{HbSBR} to Eq.~\eqref{HKerr}, we see that we also have the terms that exactly match the requirements for Kerr cat generation. However, when we compare the strength of the Kerr term to the nonlinear damping term, we find $K / \Gamma_2 = \kappa / 16 \omm$, so this term will always be small compared to the two-phonon loss rate in the sideband-resolved regime. Nonetheless, we can calculate the size of the cat state that it would generate and find
\begin{equation}
\label{betaKCom}
\beta_{\rm kc} = \sqrt{\frac{|\epst|}{K}} = \sqrt{\frac{8|\epsd| \omm^2}{\kappa g_0 |\go|}} = \sqrt{\frac{8 \omm}{\kappa}} \beta_{\rm de}.
\end{equation}
So even though the Kerr cat state generation is weaker, it attempts to make a cat state that is larger than the reservoir engineering process by a factor of $\sqrt{8 \omm / \kappa}$.

\end{document}